\documentclass[showpacs,showkeys,,preprintnumbers,amsmath,amssymb]{revtex4}
\usepackage[dvips]{graphicx}
\newcommand{\be}{\begin{equation}}
\newcommand{\ee}{\end{equation}}
\newcommand{\bea}{\begin{eqnarray}}
\newcommand{\eea}{\end{eqnarray}}
\newcommand{\nn}{\nonumber\\}
\newcommand{\oh}{\frac{1}{2}}

\newcommand{\ov}{\overline}

\newcommand{\pr}{\prime}
\newcommand{\pa}{\partial}
\newcommand{\la}{\langle}
\newcommand{\ra}{\rangle}
\def\a{\alpha}
\def\b{\beta}
\def\g{\gamma}
\def\d{\delta}
\begin{document}

\title{Self Consistent Random Phase Approximation and the
restoration of symmetries within the three-level Lipkin model}

\author{D. S. Delion}

\affiliation{National Institute of Physics and Nuclear Engineering,
 Bucharest M\u agurele, POB MG-6, Romania}

\author{P. Schuck}

\affiliation{Institut de Physique Nucl\'eaire, Orsay,
91406 - Orsay CEDEX, France}

\author{J. Dukelsky}

\affiliation{Instituto de Estructura de la Materia, CSIC, Serrano 123,
28006 Madrid, Spain}

\date{today}

\begin{abstract}
{We show that it is possible to restore the symmetry associated
with the Goldstone mode within the Self Consistent Random Phase
Approximation (SCRPA) applied to the three-level Lipkin model.
We determine one and two-body densities as very convergent
expansions in terms of the generators of the RPA basis.
We show that SCRPA excitations correspond to the heads of
some rotational bands in the exact spectrum.
It turns out that the SCRPA eigenmodes for $N=2$ coincide
with exact solutions, given by the diagonalisation procedure.}
\end{abstract}

\vskip1cm


\maketitle
\vfill\eject

\section{Introduction}
\label{sec:intro}
\setcounter{equation}{0}
\renewcommand{\theequation}{1.\arabic{equation}}

The Random Phase Approximation (RPA) is one of the most popular
microscopic approaches to describe collective excitations in interacting
many body problems \cite{Bla86,Mah81,Neg88,Fet71,Rin80}.
For example it is widely used in condensed matter physics
\cite{Mah81,Neg88,Fet71}, mainly to calculate the excitation spectrum via the
linear response function. However, RPA is equally common in nuclear physics
and other systems of finite size like cold atoms in traps
\cite{Rin80,Pet02,Pit03}, or metallic clusters \cite{Cat96}, etc.
In these various fields RPA has slightly different meaning.
We here will use it in the most general sense, like it is mostly
employed e.g. in nuclear physics, including exchange,
as it appears in linearising the time dependent Hartree-Fock equations
around equilibrium \cite{Bla86,Rin80}.
Besides its relative simplicity, both conceptually and for numerical
applications, RPA also has very desirable properties: when it is
evaluated in the Hartree-Fock (HF) basis, then the f-sum rule and
conservation laws are fulfilled, even and in fact most importantly
in cases with spontaneously broken symmetries.
For example in nuclear physics, where HF always
breaks translational invariance, the corresponding RPA restores
the symmetry in producing a Goldstone mode at zero energy \cite{Bla86,Rin80}.
The same happens when other symmetries, like rotation (deformation)
and particle number symmetries (superfluidity) are broken.
Unfortunately, when going beyond the HF-RPA scheme, the situation
with respect to symmetry restoration and conservation laws is much
less clear. Though there exists the well-known concept proposed by
Kadanoff and Baym of the $\Phi$-derivable functional \cite{Kad61},
which formally maintains all these properties,
in practice any approach beyond HF-RPA faces
serious difficulties. This stems essentially from the fact
that in this scheme going beyond RPA is equivalent to introduce
an integral kernel depending on more than one energy variable corresponding
to the s,u, and t-channels of relativistic field theory.
This then complicates the numerical task enormously.

On the other hand, in certain circumstances one has to go beyond RPA
because this approach also has a number of inherent quite severe drawbacks.
For example it is well known that RPA violates the Pauli principle
(the famous "quasi boson approximation").
This can induce strong errors and even give rise to qualitatively
wrong results, as for example in the multilevel pairing Hamiltonian
with Kramer's degeneracy, which is very much used recently for the
description of superconducting nano grains \cite{Hir02}.

A related problem which immediately appears when giving a close look to
RPA is its lack of consistency. For example in the bubble summation
one calculates the correlation energy of an electron gas \cite{Fet71}
but in the RPA-equation occupation numbers
$n^{(0)}_k=\{0,1\}$ are used which correspond to an uncorrelated (HF)
ground state. It seems obvious that, at least for the occupation numbers,
a correlated ground state should be used. The corresponding approach already
will contain RPA correlations self-consistently.
This has become increasingly popular in recent years,
and it has been called renormalised RPA (r-RPA) \cite{Har64,Row68a,Row68b}.
Lately r-RPA was widely and successfully used in describing the
double beta decay process \cite{Toi95,Suh97,Rad98} and the electronic properties
of the metal clusters \cite{Cat96}.

In fact r-RPA is an approximation to a more general extension of the RPA
which is called Self-Consistent RPA (SCRPA) \cite{Duk98}, or also
Cluster Hartree-Fock (CHF) \cite{Rop95}.

We recently had quite remarkable success with SCRPA in a number
of non trivial problems \cite{Hir02,Sto03,Kru94}.

In this paper we will specifically investigate the properties of SCRPA
with respect to conservation laws and restoration of symmetries,
the fulfilement of which is,
as already mentioned, one of the outstanding aspects of
standard RPA theory. It will be shown that for a wide class of symmetries
SCRPA can be formulated in such a way that it also {\it maintains}
these properties of standard RPA.
This is to be considered as a strong advantage over other
extensions of RPA because SCRPA, though much more demanding
numerically than standard RPA, leads to an (nonlinear) equation of
the Schr\"odinger type which seems to be accessible for a
numerical solution.
The restoration of symmetries within SCRPA is directly linked
to the consideration of the so-called "scattering" terms
and we will discuss their significance in detail.

Here we will demonstrate those properties of SCRPA in an exactly solvable
three level Lipkin model \cite{Li70,Mes71}, which can also be interpreted as an
interacting spin problem with two sites. Recently
this model was investigated within the multistep
variational approach \cite{Sam99} and the r-RPA formalism \cite{Gra02}.
This model has the property that, as swon in Ref. \cite{Hag00},
once the two upper levels become degenerate, a continuously broken
symmetry develops beyond a certain critical coupling which can be considered
as a deformed state with broken rotational symmetry.
We will show that SCRPA, as standard RPA, will develop a zero mode
(Goldstone mode) beyond a critical coupling, thus exactly restoring
the original invariance of the Hamiltonian.

We also will see that SCRPA yields much improved solutions
which we will compare with the ones obtained
from an exact diagonalization. For the two particle case SCRPA
even reproduces the exact results.
In detail the paper is organized as follows.
In Section 2 we give a short description of the SCRPA
and emphasise the main differences with respect to r-RPA.
We outline here and in the Appendix B the method to determine
the one and two-body densities.
In Section 3 we give some numerical examples concerning especially
the Goldstone mode.
In Appendix A we describe a very efficient procedure to solve
SCRPA as a system of coupled equations, based on r-RPA as a first step.
In the last Section we draw our conclusions.

\section{Theoretical background}
\label{sec:theor}
\setcounter{equation}{0}
\renewcommand{\theequation}{2.\arabic{equation}}

Let us consider a many body system of fermions characterized by $n+1$
single particle levels, labelled $\a=0,1,2,...,n$.
We define the "quadrupole-like" basis operators of this model
\be
\label{Kab}
K_{\a\b}\equiv\sum_{\mu=1}^Nc^{\dag}_{\a\mu}c_{\b\mu}~,
\ee
where $c^{\dag}_{\a\mu}$ denotes the creation operator 
of a fermion on some $\a$-th level.
We suppose the levels degenerate, the total number of projections
$\mu$ being $N=2\Omega$.
The operators (\ref{Kab}) satisfy simple commutation rules, namely
\be
\label{comut}
\left[K_{\a\b},K_{\g\d}\right]=\delta_{\b\g}K_{\a\d}-\delta_{\a\d}K_{\g\b}~.
\ee
We will consider the following general Hamiltonian built on these operators
\bea
\label{Hamilt0}
H=\sum_{\a=0}^n\epsilon_{\a}K_{\a\a}+\sum_{\a\b\g\d=0}^n
V_{\a\b\g\d}K_{\a\b}K_{\g\d}~.
\eea
We have chosen in our numerical application the three-level Lipkin model,
with $n=2$, corresponding to the SU(3) algebra.
This model has been widely used in order to test different
many-body approximations \cite{Li70,Mes71,Sam99,Gra02,Hag00}.
In analysing this model we have used a particular form of the
Hamiltonian (\ref{Hamilt0}), namely
\bea
\label{Hamilt}
H=\sum_{\a=0}^2\epsilon_{\a}K_{\a\a}
-\frac{V}{2}\sum_{\a=1}^2(K_{\a 0}K_{\a 0}+K_{0\a}K_{0\a})
~.
\eea
It is a particular case of the Hamiltonian used in Ref. \cite{Hol74}
and it is analyzed in detail in Ref. \cite{Hag00} in connection
with a continuously broken symmetry which appears, as already mentioned,
when $\epsilon_1=\epsilon_2$.

We will analyze the SCRPA in order to obtain the excitation energies
and we will compare the results with the r-RPA scheme as well as with the
exact results, obtained by the diagonalization procedure.

We recall that both SCRPA and r-RPA schemes involve several common steps,
namely \cite{Duk90}

A) One finds the minimum of generalised mean field equations
coupled to the RPA fluctuations.

B) One determines the RPA eigenstates.

C) The RPA amplitudes are used to compute one and two-body correlated
densities.
At this step r-RPA supposes a factorisation for the
two-body densities, while SCRPA uses a fully correlated approach.

The amplitudes are used to iterate the steps A, B and C until
the convergency is achieved.

In the following we will shortly describe this procedure.

\subsection{Mean field}
\label{subsec:theora}

As a first step in both SCRPA and r-RPA schemes we set up
the generalised mean field equations.
To this purpose let us introduce fermion creation operators in
a general single particle basis as follows
\be
\label{HF1}
a^{\dag}_{k\mu}=\sum_{\a=0}^nC_{k\a}c^{\dag}_{\a\mu}~,
\ee
where the supposed real transformation matrix is orthogonal.
The unity matrix for $C_{k\alpha}$ corresponds to a "spherical" minimum,
while a non-diagonal transformation corresponds to a "deformed" one.
According to Ref. \cite{Hag00} for the three-level Lipkin model
the above real matrix can be written as a product of two rotations, i.e.
\bea
\label{matrix}
C_{k\a} &=&
\left(\begin{matrix}
 {\rm cos}\phi & {\rm sin}\phi & 0 \cr
-{\rm sin}\phi & {\rm cos}\phi & 0 \cr
     0     &     0     & 1 \cr
\end{matrix}\right)
\left(\begin{matrix}
     1     &     0      &     0     \cr
     0     &  {\rm cos}\psi  &  {\rm sin}\psi \cr
     0     & -{\rm sin}\psi  &  {\rm cos}\psi \cr
\end{matrix}\right)
\nn
&=&\left(\begin{matrix}
 {\rm cos}\phi & {\rm sin}\phi~{\rm cos}\psi & {\rm sin}\phi~{\rm sin}\psi \cr
-{\rm sin}\phi & {\rm cos}\phi~{\rm cos}\psi & {\rm cos}\phi~{\rm sin}\psi \cr
     0     &   -{\rm sin}\psi    &     {\rm cos}\psi     \cr
\end{matrix}\right)~.
\eea
The basic operators (\ref{Kab}) are replaced in this new basis by
similar combinations of the new operators (\ref{HF1})
\bea
\label{A}
A_{ij}&=&\sum_{\mu=1}^N a^{\dag}_{i\mu}a_{j\mu}~.
\eea
By using the inverse transformation of (\ref{HF1}) the Hamiltonian
(\ref{Hamilt0}) takes the following form in the new basis
\be
\label{HamHF}
H=\sum_{ij=0}^n
F_{ij}A_{ij}
+ \sum_{ijkl=0}^nG_{ijkl}A_{ij}A_{kl}~,
\ee
where the coefficients are given respectively by
\bea
\label{FG}
F_{ij}&=&\sum_{\a=0}^n \epsilon_{\a} C_{i\a}C_{j\a}~,
\nn
G_{ijkl}&=&\sum_{\a\b\g\d=0}^n V_{\a\b\g\d}
C_{i\a}C_{j\b}C_{k\g}C_{l\d}~.
\eea

The expectation value of the Hamiltonian (\ref{HamHF}) with the correlated
vacuum $|0\rangle$ has the form
\be
\label{expHam}
{\cal
H}\equiv\langle 0|H|0\rangle
=\sum_{ij=0}^nF_{ij}\langle ij\rangle
+\sum_{ijkl=0}^nG_{ijkl}\langle ijkl\rangle~,
\ee
where we introduced one and two-body densities as follows
\bea
\langle ij\rangle&\equiv&\langle 0|A_{ij}|0\rangle ~,
\nn
\langle ijkl\rangle&\equiv&\langle0|A_{ij}A_{kl}|0\rangle~.
\eea
In order to derive the Mean Field (MF) equations let us now consider
the following restricted functional
\be
{\cal H}^{\pr}={\cal H}-\sum_{\a}E_{\a}
\langle \a\a\rangle
\sum_{k}C_{k\a}C_{k\a}~.
\ee
We below will show that the terms with non-diagonal densities
$\la\alpha\beta\ra$
vanish because they contain generators of the RPA basis, which can be
inverted in terms of SCRPA annihilation phonons acting on the vacuum state.
The extremum condition
\be
\label{mfeq}
\frac{\pa {\cal H}^{\pr}}{\pa C_{n\a}}=0
\ee
leads to the following system of equations
\bea
\label{HFeq1}
\epsilon_{\a}\langle nn\rangle C_{n\a}
+\sum_{jkl}\sum_{\b\g\d}
(njkl;\a\b\g\d)C_{j\b}C_{k\g} C_{l\d}
=E_{n}\langle nn\rangle C_{n\a}~,
\eea
where we introduced as short-hand notation
\bea
\label{abcd}
(njkl;\a\b\g\d)\equiv
\frac{1}{2}[
\langle njkl\rangle V_{\a\b\g\d}+
\langle jnkl\rangle V_{\b\a\g\d}+
\langle kjnl\rangle V_{\g\b\a\d}+
\langle ljkn\rangle V_{\d\b\g\a}]~.
\eea
By inserting a formal summation $\sum_{\mu}\delta_{n\mu}$
and using the orthonormality condition for the MF amplitudes
one obtains an eigenvalue problem
\be
\label{HFeq2}
\sum_{m} H_{nm}C_{m\a} = E_{\alpha}
\langle\alpha\alpha\rangle C_{n\a}~.
\ee
We solve this system to determine the eigenvalues
$E_{\alpha}$ and eigenvectors $C_{m\a}$ for $\alpha=0,1,...,N$.
In case we evaluate expectation values (\ref{expHam}) using the
standard HF vacuum one obtains what we shall call the "static minimum".
If one uses the SCRPA correlated vacuum the result is
called "dynamic minimum" (coupled with RPA fluctuations).
In both cases the system (\ref{HFeq2}) contains nonlinear relations,
because the matrix $H_{nm}$
\bea
\label{Hnm}
H_{nm}
&=&\langle nn\rangle\sum_{\mu}\epsilon_{\mu}C_{n\mu}C_{m\mu}
+\sum_{jkl}\sum_{\mu\beta\gamma\delta}
(njkl;\mu\b\g\d)C_{m\mu}C_{j\b}C_{k\g}C_{l\d}
\eea
is written in terms of the eigenvectors and therefore
it should be solved iteratively.
For the "dynamic minimum" the iterative process is more complex
because it involves the correlated vacuum, i.e. RPA amplitudes.
As an initial solution we can consider the standard HF problem,
leading to a decoupled form for the two-body matrix elements
\bea
\label{dec}
\langle ijkl\rangle\approx
\langle ii\rangle\langle kk\rangle
\delta_{ij}\delta_{kl}+
\langle ii\rangle (1-\langle kk\rangle/N)
\delta_{il}\delta_{jk}~.
\eea
With $\langle ii\rangle\equiv\rho_i^{(0)}=\{1,0\}$,
one obtains directly from Eq. (\ref{HFeq1}) the standard system
of HF equations
\bea
\label{HF0}
\sum_{m} H^{(0)}_{nm}C_{m\a}
=E_{\a}^{HF}C_{m\a}~,
\eea
where $H^{(0)}_{nm}$ are the elements of the HF Hamiltonian,
which we will give for a particular interaction considered
in our numerical application.
However, in general, it is not necessary to consider the 
$\langle kk\rangle$ in (\ref{dec}) as uncorrelated. With
correlations included Eq. (\ref{dec}) will lead to the
so-called renormalised RPA (r-RPA) (see below).

\subsection{SCRPA equations}
\label{subsec:theorb}

The procedure to derive the system of RPA equations is given in many
textbooks, see e.g. \cite{Bla86,Rin80}.
Here we will shortly recall the main steps in order to
introduce the necessary notations.
To this purpose we use the equation of motion technique.
The SCRPA creation operators are defined by
the following superposition of the basic pair generators
\be
\label{phonon}
Q^{\dag}_{\nu}=\sum_{m>i}
(X_{mi}^{\nu}\delta Q^{\dag}_{mi}-Y_{mi}^{\nu}\delta Q_{mi})~.
\ee
These operators are written in the MF basis as follows
\be
\label{gen}
\delta Q^{\dag}_{mi}=N^{-1/2}_{mi}A_{mi},~~~
\delta Q_{mi}=\delta Q^{\dag}_{im}=N^{-1/2}_{mi}A_{im}~.
\ee
and define the excited states
\bea
|\nu\rangle=Q^{\dag}_{\nu}|0\rangle~.
\eea
The RPA annihilation operators define a vacuum state, i.e.
\bea
\label{annih}
Q_{\nu}|0\rangle=0~.
\eea

It is important to stress that we used the letters $m$ and
$i$ not for labelling holes and particles, but only
to remember the condition $m>i$ as written in (\ref{phonon}).
In this way we include in the basis all possible particle-hole (ph)
and also the so called "scattering"
elements with particle-particle (pp) and hole-hole (hh) configurations.
This type of amplitudes is absent in standard RPA and we will
have to discuss their significance in the course of the present paper.
It has been recently shown that the inclusion of scattering
terms allows to fulfil the energy weighted sum rule
in the renormalised RPA (r-RPA) \cite{Gra01}
and they may therefore be of importance.

The normalisation factor in (\ref{gen}) is given by the following mean value
on the selfconsistent vacuum
\be
\label{norm}
\langle 0|[A_{im},A_{nj}]|0\rangle
=\delta_{mn}\delta_{ij}
(\langle ii\rangle-\langle mm\rangle)\equiv\delta_{mn}\delta_{ij}N_{mi}~.
\ee
We then evaluate the mean value of the double commutator
between the Hamiltonian and the phonon operator (\ref{phonon}) \cite{Duk90}.
One obtains in this way the SCRPA system of equations
\bea
\label{RPAsys}
\left(\begin{matrix}
{\cal A} & {\cal B} \cr -{\cal B}^* & -{\cal A}^* \cr
\end{matrix}\right)
\left(\begin{matrix}X^{\nu} \cr Y^{\nu} \cr\end{matrix}\right)=\omega_{\nu}
\left(\begin{matrix}X^{\nu} \cr Y^{\nu} \cr\end{matrix}\right)~,
\eea
where the matrices are given by the well-known relations
\bea
\label{AB}
{\cal A}_{mi,nj}&=&
 \langle 0|\left[\delta Q_{mi},\left[H,\delta Q^{\dag}_{nj}\right]\right]
 |0\rangle
\nn
{\cal B}_{mi,nj}&=&
-\langle 0|\left[\delta Q_{mi},\left[H,\delta Q_{nj}\right]\right]
 |0\rangle =-{\cal A}_{mi,jn}~.
\eea
Additionally we have the generalised mean field equations (\ref{mfeq}),
which with (\ref{annih}) can also be written as
\bea
\langle 0|[H,Q_{\nu}] |0\rangle=
\langle 0|[H,Q_{\nu}^{\dag}] |0\rangle=
\langle 0|[H,\delta Q_{\nu}^{\dag}] |0\rangle=0~.
\eea
The amplitudes satisfy usual normalisation and completeness conditions
\bea
\sum_{m>i}\left(
X^{\nu}_{mi}X^{\mu}_{mi}-Y^{\nu}_{mi}X^{\mu}_{mi}
\right)&=&\delta_{\nu,\mu}~,
\nn
\sum_{\nu}\left(
X^{\nu}_{mi}X^{\nu}_{nj}-Y^{\nu}_{mi}X^{\nu}_{nj}
\right)&=&\delta_{mi,nj}~.
\eea

For the general Hamiltonian (\ref{HamHF}) one obtains the following
form of the SCRPA matrices
\bea
\label{ARPA}
a_{mi,nj}&\equiv& N_{mi}^{1/2}N_{nj}^{1/2}{\cal A}_{mi,nj}
\nn
&=&{\oh}(\delta_{ij}F_{mn}-\delta_{mn}F_{ji})
(\langle ii\rangle+\langle jj\rangle-\langle mm\rangle-\langle nn\rangle)
\nn
&+&\sum_{ab}
(G_{mabn}\langle iabj\rangle
+G_{abmn}\langle abij\rangle)_S
+\sum_{ab}
(G_{aijb}\langle amnb\rangle
+G_{abji}\langle abnm\rangle)_S
\nn
&-&\sum_{ab}
(G_{majb}\langle ianb\rangle
+G_{aibn}\langle ambj\rangle)_S
~,
\nn
&-&{\oh}\delta_{ij}\sum_{abc}
(G_{abcn}\langle abcm\rangle
+G_{abmd}\langle abnc\rangle)_S
\nn
&-&
{\oh}\delta_{mn}\sum_{abc}
(G_{abci}\langle abcj\rangle
+G_{abjc}\langle abic\rangle)_S~,
\nn
b_{mi,nj}&=&-a_{mi,jn}~,
\eea
where we used the following short-hand notation
\bea
\label{Gsym}
(G_{ijkl}\langle abcd\rangle)_S
\equiv
G_{ijkl}\langle abcd\rangle+
G_{klij}\langle cdab\rangle~.
\eea

The r-RPA differs from the SCRPA procedure in the way to determine
the two-body densities.
By using the factorised ansatz (\ref{dec}) one obtains the r-RPA matrices
\bea
\label{ABren}
{\cal A}^{(0)}_{mi,nj}&=&
{\oh}(\delta_{ij}H^{(0)}_{mn}-\delta_{mn}H^{(0)}_{ji})
(N^{1/2}_{mi}N^{-1/2}_{nj}+N^{-1/2}_{mi}N^{1/2}_{nj})
\nn
&+&\left(\ov{G}_{mijn}-\frac{1}{N}\ov{G}_{jimn}\right)N^{1/2}_{mi}N^{1/2}_{nj}
\nn
{\cal B }^{(0)}_{mi,nj}&=&
{\oh}(\delta_{mj}H^{(0)}_{ni}-\delta_{in}H^{(0)}_{mj})
(N^{1/2}_{mi}N^{-1/2}_{nj}-N^{-1/2}_{mi}N^{1/2}_{nj})
\nn
&+&\left(\ov{G}_{minj}-\frac{1}{N}\ov{G}_{nimj}\right)N^{1/2}_{mi}N^{1/2}_{nj}~,
\eea
where we used the following notations
\bea
\ov{G}_{minj}=G_{minj}+G_{njmi}~,
\eea
and
\bea
\label{HFeq}
H^{(0)}_{mn}\equiv F_{mn}+
\sum_{a}\left[G_{maan}+
\left(\ov{G}_{aa mn}-\frac{1}{N}\ov{G}_{maan}\right)
\langle aa\rangle\right]~.
\eea
With the help of the MF equation (\ref{HF0})
one obtains that $H^{(0)}_{mn}$ is diagonal, i.e.
\bea
\label{emn}
H^{(0)}_{mn}=\delta_{mn} E_n~.
\eea
We should stress that the occupation numbers appearing in
(\ref{HFeq}) are calculated, as will be explained later,
with the correlated vacuum and therefore
(\ref{emn}) is {\it not} equivalent to standard HF,
even within the r-RPA scheme.

For r-RPA amplitudes with respect to the initial basis $A_{mi}$
\bea
\label{redamp}
x^{\nu}_{0,mi}\equiv N_{mi}^{-1/2}X^{\nu}_{0,mi}~,~~~
y^{\nu}_{0,mi}\equiv N_{mi}^{-1/2}Y^{\nu}_{0,mi}~,
\eea
one obtains expressions with no phase space factors concerning the indices
$m,i$, i.e. there appear no occupation numbers
\bea
\label{redamp1}
x^{\nu}_{0,mi}&=&\frac{1}{\omega^{(0)}_{\nu}-(E_m-E_i)}
\sum_{nj}\left[
\left(\ov{G}_{mijn}-\frac{1}{N}\ov{G}_{jimn}\right)\ov{X}^{\nu}_{0,nj}+
\left(\ov{G}_{minj}-\frac{1}{N}\ov{G}_{nimj}\right)\ov{Y}^{\nu}_{0,nj}
\right]~,
\nn
y^{\nu}_{0,mi}&=&\frac{-1}{\omega^{(0)}_{\nu}+(E_m-E_i)}
\sum_{nj}\left[
\left(\ov{G}_{mijn}-\frac{1}{N}\ov{G}_{jimn}\right)\ov{Y}^{\nu}_{0,nj}+
\left(\ov{G}_{minj}-\frac{1}{N}\ov{G}_{nimj}\right)\ov{X}^{\nu}_{0,nj}
\right]~,
\nn
\eea
where we defined the new amplitudes
\bea
\label{redamp2}
\ov{X}^{\nu}_{0,mi}\equiv N_{mi} x^{\nu}_{0,mi}~,~~~
\ov{Y}^{\nu}_{0,mi}\equiv N_{mi} y^{\nu}_{0,mi}~.
\eea
We mention here that the r-RPA amplitudes defined by Eqs. (\ref{redamp})
and (\ref{redamp2}) are very important in solving numerically the
SCRPA system of equations using the two-step method described
in the Appendix A.

\subsection{SCRPA densities}
\label{subsec:theorc}

In order to estimate the expectation values of
the one-body density operators within the SCRPA
\be
\label{dens}
\hat{\rho}_i\equiv\sum_{\mu} a^{\dag}_{i\mu}a_{i\mu},~~i=0,1,2~,
\ee
we need a special procedure. In the following we propose a method
which is more convergent than the standard expansion in terms
of phonon operators (\ref{phonon}).
To this purpose let us introduce the shell model basis,
but in the MF representation and which do not necessarily
correspond to the HF basis, as we see in (\ref{HFeq2})
\be
\label{Abasis}
|n_1n_2\rangle\equiv
{\cal N}^{-1/2}_{n_1n_2}A^{n_1}_{10}A^{n_2}_{20}|MF \rangle,~~
0\leq n_1+n_2\leq N~,
\ee
where $|MF\rangle$ corresponds to a Slater determinant built with
operators (\ref{HF1}).
Within this complete basis we can expand a general combination
of one-body densities 
\bea
\label{expans}
\hat{\rho}_0^{k_0}\hat{\rho}_1^{k_1}\hat{\rho}_2^{k_2}
&=&\sum_{n_1n_1=0}^{N}b_{n_1n_2}(k_0k_1k_2)|n_1n_2\rangle
\langle n_2n_1|
\nn
&=&\sum_{n_1n_1=0}^{N}c_{n_1n_2}(k_0k_1k_2)
A_{10}^{n_1}A_{20}^{n_2}A_{02}^{n_2}A_{01}^{n_1}~,
\eea
where $k_0,k_1,k_2$ are some given exponents. The connection between
the $b$ and $c$ coefficients is given by iterating the resolution
of the unity.
This method is used in Ref. \cite{Sch71} and \cite{Feu00} .
For us it is important to find the coefficients $c_{n_1n_2}(k_0k_1k_2)$
and we will use a more direct technique.
To this purpose we need the expectation values
of the operators (\ref{expans}) on the correlated vacuum $|0\rangle$.
They are calculated by using the inversion of (\ref{phonon}).
One obtains
\bea
\label{Ainvers}
A_{mi}&=&
N_{mi}^{1/2}
\sum_{\nu} (X_{mi}^{\nu}Q^{\dag}_{\nu}+Y_{mi}^{\nu}Q_{\nu})~,
\nn
A_{im}&=&
N_{mi}^{1/2}
\sum_{\nu} (X_{mi}^{\nu}Q_{\nu}+Y_{mi}^{\nu}Q^{\dag}_{\nu})~,
\eea
where the normalisation factor $N_{mi}$ is given by (\ref{norm}).
For instance in the case of products of two operators
one obtains the following relations for $m>i,n>j$
\bea
\label{AA}
\langle 0|A_{mi}A_{nj}|0\rangle&=&
N^{1/2}_{mi}N^{1/2}_{nj}\sum_{\nu}Y^{\nu}_{mi}X^{\nu}_{nj}
\nn
\langle 0|A_{im}A_{jn}|0\rangle&=&
N^{1/2}_{mi}N^{1/2}_{nj}\sum_{\nu}X^{\nu}_{mi}Y^{\nu}_{nj}
\nn
\langle 0|A_{mi}A_{jn}|0\rangle &=&
N^{1/2}_{mi}N^{1/2}_{nj}\sum_{\nu}Y^{\nu}_{mi}Y^{\nu}_{nj}
\nn
\langle 0|A_{im}A_{nj}|0\rangle&=&
N^{1/2}_{mi}N^{1/2}_{nj}\sum_{\nu}X^{\nu}_{mi}X^{\nu}_{nj}~.
\eea
We mention that these relations can be directly used to express
some of the SCRPA matrix elements in terms of RPA amplitudes.
One finally obtains from (\ref{expans}) a nonlinear system of
equations, determining the normalisation factors $N_{10},N_{20}$.
It is given in the Appendix B in terms of expansion coefficients
defined by Eq. (\ref{expans}).
These coefficients are fast decreasing with increasing $n_1+n_2$.
One can see from the Table 2 that already the linear approximation,
i.e. $n_1+n_2\leq 1$, ensures an accuracy of the order $10^{-2}$.
In this case one also directly obtains the one-body densities
\bea
\label{denlin}
\langle 0|\hat{\rho}_{m}|0\rangle&=&
\left[ y_{mm}+\frac{y_{11}y_{22}}{N} \right]
\left[ 1+\frac{2}{N}(y_{11}+y_{22})+
\frac{3}{N^2}y_{11}y_{22} \right]^{-1},~~m=1,2~,
\nn
\langle 0|\hat{\rho}_{0}|0\rangle&=&
N-\langle 0|\hat{\rho}_{1}|0\rangle-\langle 0|\hat{\rho}_{2}|0\rangle~,
\eea
where
\bea
y_{mn}=\sum_{\nu}Y^{\nu}_{m0}Y^{\nu}_{n0}~.
\eea

Based on this result one may hope that also for a more realistic $n+1$-level
model one can expand the product of densities in the $ph$
"shell-model" basis, i.e.
\bea
\label{Abasisn}
|n_1n_2...n_n\rangle\equiv {\cal N}^{-1/2}_{n_1n_2...n_n}
A^{n_1}_{10}A^{n_2}_{20}...A^{n_n}_{n0}|0\rangle,~~
0\leq n_1+n_2+...+n_n\leq 1~.
\eea

\section{Numerical application}
\label{sec:numeric}
\setcounter{equation}{0}
\renewcommand{\theequation}{3.\arabic{equation}}

\subsection{Hartree-Fock mean field}
\label{subsec:numa}

In order to apply the SCRPA and r-RPA procedures it is convenient
to start iterations using standard HF and RPA solutions.
To this purpose let us first evaluate the expectation values of the one
and two-body operators on the standard HF vacuum
\bea
\label{HFdens}
\langle\alpha\beta\rangle&=&\delta_{\alpha\beta}\delta_{\alpha 0}N~,
\nn
\langle\alpha\beta\gamma\delta\rangle&=&
\delta_{\alpha 0}\delta_{\delta 0}\delta_{\beta\gamma}N
\left[1+\delta_{\beta 0}(N-1)\right]~.
\eea
We stress that the two-body density has the same form as the factorised
ansatz (\ref{dec}).
The expectation value of the Hamiltonian (\ref{Hamilt}) then becomes
\bea
\langle H\rangle&=&F_{00}N+\sum_{\alpha}
G_{0\alpha\alpha 0}N+G_{0000}N(N-1)
\nn
&=&N\epsilon [
e_0 {\rm cos}^2\phi + e_1 {\rm sin}^2\phi {\rm cos}^2\psi +
 e_2 {\rm sin}^2\phi {\rm sin}^2\psi
\nn
&-&\chi {\rm sin}^2\phi
 {\rm cos}^2\phi
 ]~,
\eea
where we introduced the following dimensionless notations
\bea
e_k=\frac{\epsilon_k}{\epsilon},~~~\chi=\frac{V(N-1)}{\epsilon}~.
\eea

The Hamiltonian (\ref{Hamilt}) has 
two kinds of HF minima, namely a spherical minimum and a deformed one
\bea
\label{HFstat}
&1)&~~~\phi=0,~~~\psi=0~,~~~\chi<e_1-e_0~,
\nn
&2)&~~~{\rm cos}~2\phi=\frac{e_1-e_0}{\chi},~~~\psi=0,~~~\chi>e_1-e_0~.
\eea
Moreover, our calculations have shown that for any MF minimum one
obtains $\psi=0$, defined by the parametrisation (\ref{matrix}),
independent of which kind of vacuum (correlated or not)
we use to estimate the expectation values.

\subsection{Standard RPA}
\label{subsec:numb}

For the above mentioned minima we obtain that the standard
RPA matrix elements have very simple expressions.
By using the following short-hand notation for the index pairs
\bea
10 \rightarrow 1,~20 \rightarrow 2,~21 \rightarrow 3
\eea
one obtains \cite{Hag00}
\bea
\label{RPA}
{\cal A}_{11}&=&\epsilon
(e_1-e_0){\rm cos}~2\phi+\epsilon\frac{3}{2}\chi {\rm sin}^2~2\phi~,
\nn
{\cal A}_{22}&=&\epsilon
(e_2-e_0 {\rm cos}^2\phi-e_1 {\rm sin}^2\phi)+
\epsilon{\oh}\chi {\rm sin}^22\phi~,
\nn
{\cal A}_{12}&=&{\cal A}_{21}=0~,
\nn
{\cal B}_{11}&=&-\epsilon\chi({\rm cos}^4\phi+{\rm sin}^4\phi)~,
\nn
{\cal B}_{22}&=&-\epsilon\chi{\rm cos}^2\phi~,
\nn
{\cal B}_{12}&=&{\cal B}_{21}=0~.
\eea
Therefore the ${\cal A}$ and ${\cal B}$ RPA matrices are diagonal.
The RPA frequencies are easy to estimate
\bea
\label{RPAfre}
\omega_k^2= {\cal A}_{kk}^2-{\cal B}_{kk}^2~,~~~k=1,2~,
\eea
and the amplitudes become
\bea
\label{RPAamp}
\left(\begin{matrix} X_k^{\nu}\cr Y_k^{\nu} \end{matrix} \right)=
\frac{1}{\sqrt{2}}
\left[\frac{{\cal A}_{kk}}{\omega_k} \pm 1\right]^{1/2} \delta_{k\nu}
\eea
We fix the origin of the particle spectrum with $e_0=0$.
Then for a spherical vacuum $\phi=0$ the RPA energies are given by
\bea
\label{RPAfre0}
\omega_{\nu}=\epsilon_{\nu}
\left[ 1-\left(\frac{\chi}{e_{\nu}}\right)^2\right]^{1/2}~,
~~~\nu=1,2~,
\eea
with the corresponding RPA amplitudes
\bea
\label{RPAamp0}
\left(\begin{matrix} X_k^{\nu}\cr Y_k^{\nu} \end{matrix} \right)=
\frac{1}{\sqrt{2}}
\left[\frac{\epsilon_k}{\omega_k} \pm 1\right]^{1/2} \delta_{k\nu}
~.
\eea

As it was shown in Ref. \cite{Hag00}, if the upper single particle
levels are degenerate, i.e. $\Delta\epsilon\equiv\epsilon_2-\epsilon_1=0$,
for the values of the strength $\chi>e_1$, in the "deformed region",
i.e. with $\phi\neq 0$ given by HF minimum, one obtains a Goldstone mode.
In this case by considering $e_1=1$ one obtains for the excitation energies
\bea
\label{RPAfre1}
\omega_1&=&\epsilon\sqrt{2(\chi^2-1)}~,
\nn
\omega_2&=&0~.
\eea
Indeed, in Fig. 1 by dashed lines are plotted standard RPA
excitation energies versus the strength parameter $\chi$, by considering
the number of particles $N=10$.
For $\chi<1$ (spherical region) the eigenvalues $\omega_1,\omega_2$
are degenerate, while for $\chi>1$ (deformed region, $\phi\neq 0$)
the second mode $\omega_2$ has a vanishing value, i.e.
it corresponds to the Goldstone mode.
In this case the amplitude of the non-vanishing mode are given by
\bea
\left(\begin{matrix} X_1^1\cr Y_1^1 \end{matrix} \right)=
\frac{1}{\sqrt{2}}
\left(\frac{3\chi-1/\chi}{2\sqrt{2(\chi^2-1)}}\pm 1\right)~,~~~
X_2^1=Y_2^1=0~.
\eea
The corresponding amplitudes for the Goldstone mode are given by
\bea
\label{Gold}
X^{2}_{1}=Y^{2}_{1}=0~,~~~
X^{2}_{2}=Y^{2}_{2}\rightarrow\infty~.
\eea
A similar plot for $N=20$ is given in Fig. 2.

\subsection{SCRPA for N=2}
\label{subsec:numc}

It is possible to find the exact eigenstates
of the three-level Lipkin Hamiltonian (\ref{Hamilt})
by diagonalizing it in the following normalized basis
\bea
\label{basis}
|n_1n_2\ra=\sqrt{\frac{(N-n_1-n_2)!}{N!n_1!n_2!}}
K^{n_1}_{10}K^{n_2}_{20}|HF\ra~.
\eea
In this paragraph we will pay special attention to the
spherical case $\phi=0$, with $N=2$, i.e. the two-particle case.
The above diagonalisation basis contains 6 elements with $n_1+n_2\leq 2$.
The calculation shows that the non-vanishing components of the ground state
are $|n_1n_2\ra=|00\ra,|20\ra,|02\ra$. The first two excited states
contain only $|10\ra$ and $|01\ra$ components, respectively.
This fact suggests that we can consider the two component phonon,
with components $(10)\rightarrow 1, (20)\rightarrow 2$.
The corresponding SCRPA vacuum state has the following form
\bea
\label{vacuum}
|0\ra=\left(a_0+\frac{a_1}{2}K_{10}^2+\frac{a_2}{2}K_{20}^2
\right)|MF\ra~.
\eea
By using the action of the annihilation operator on the vacuum
(\ref{annih}) one obtains for the coefficients
\bea
a_0&=&\left[1+\left(YX^{-1}\right)_{11}^2+\left(YX^{-1}\right)_{22}^2
\right]^{-1/2}~,
\nn
a_1&=&a_0\left(YX^{-1}\right)_{11}~,
\nn
a_2&=&a_0\left(YX^{-1}\right)_{22}~.
\eea
The one-body densities and their products can be expressed in
terms of the vacuum amplitudes as follows
\bea
\label{d1}
\la 0 | \hat{\rho}_k| 0\ra&=&2a_k^2~,~~~
\la 0 | \hat{\rho}_k^2| 0\ra=4a_k^2~,~~~k=0,1,2~,
\nn
\la 0 | \hat{\rho}_0\hat{\rho}_1| 0\ra&=&
\la 0 | \hat{\rho}_0\hat{\rho}_2| 0\ra=
\la 0 | \hat{\rho}_1\hat{\rho}_2| 0\ra=0~.
\eea
The SCRPA matrix elements of the Hamiltonian (\ref{Hamilt})
can be directly evaluated by computing double commutators expectation values
with respect to the vacuum. It turns out that they are diagonal
and therefore the two eigenstates are decoupled.
This allows us to express in a simple way the coefficients entering
the vaccuum state
\bea
z_k\equiv\left(YX^{-1}\right)_{kk}=\frac{Y_k}{X_{k}}~,~~~k=1,2~.
\eea
The SCRPA matrix elements have the following form
\bea
{\cal A}_{11}&=&\epsilon
\left[e_1-e_0+\chi\frac{2z_1+z_2}{1-z_1^2}\right]~,
\nn
{\cal A}_{22}&=&\epsilon
\left[e_2-e_0+\chi\frac{2z_2+z_1}{1-z_2^2}\right]~,
\nn
{\cal A}_{12}&=&{\cal A}_{21}=0~,
\nn
{\cal B}_{11}&=&-\epsilon\chi\frac{1+z_1^2+z_1z_2}{1-z_1^2}~,
\nn
{\cal B}_{22}&=&-\epsilon\chi\frac{1+z_2^2+z_2z_1}{1-z_2^2}~,
\nn
{\cal B}_{12}&=&{\cal B}_{21}=0~.
\eea
The matrix elements of the Hamiltonian with respect to the
above vacuum give the ground state energy as follows
\bea
E_0=2\epsilon\frac{e_1z_1^2+e_2z_2^2-\chi(z_1+z_2)}{1+z_1^2+z_2^2}~.
\eea
The SCRPA equations lead to the following nonlinear system
of equations for $z_1,z_2$
\bea
\frac{\chi}{2e_1}
\left[z_1^4-2z_1^2+\left(z_1^3-z_1\right)z_2+1\right]
+z_1^3-z_1&=&0~,
\nn
\frac{\chi}{2e_2}
\left[z_2^4-2z_2^2+\left(z_2^3-z_2\right)z_1+1\right]
+z_2^3-z_2&=&0~.
\eea
We solved this system by using the Newton procedure with a precision
of $10^{-8}$.
The results for both eigenvalues and amplitudes coincide with the exact
values using the diagonalisation procedure, independent of the used
coupling constant $\chi$ and single particle energies.
In Table 1 we give the coefficients $a_k,k=0,1,2$, the ground state
energy and the excitation energies $\omega_k,k=1,2$ for
$\chi=5,~e_0=0,~e_1=1,~e_2=2,~\epsilon=1~MeV$.
The two lines correspond both to the exact
and SCRPA values and they fully coincide.
Therefore it is possible to construct the SCRPA ground state
and excitations, which actually coincide with the exact solution.
The inclusion of the third component $(21)$ in the structure
of the vacuum state for $N=2$ gives a vanishing coefficient
and therefore it is not necessary to consider it in this case.

\begin{center}
{\bf Table 1}
\vskip5mm
{\it 
The vacuum coefficients $a_k,k=0,1,2$ in (\ref{vacuum}),
the ground state energy and the excitation energies $\omega_k,k=1,2$ at
$\chi=5,~\epsilon_n=n~MeV,~n=0,1,2$ for the exact and SCRPA values.

}
\vskip5mm
\begin{tabular}{|c|c|c|c|c|c|c|}
\hline
 & $a_0$ & $a_1$ & $a_2$ & $E_0$  & $\omega_1$ & $\omega_2$ \cr
\hline
exact~&~0.772576~&~0.496911~&~0.395228~&~-5.773794~&~6.773794~&~7.773794~\cr
SCRPA~&~0.772576~&~0.496911~&~0.395228~&~-5.773794~&~6.773794~&~7.773794~\cr
\hline
\end{tabular}
\end{center}
\vskip5mm

For $N>2$ the equation (\ref{annih}) leads to a system where
the number of conditions exceeds the number of coefficients.
In order to find the SCRPA ground state by satisfying
the condition (\ref{annih}) it is necessary to enlarge
the basis, by considering e.g. for $N=4$ two particle-two hole(2p-2h)
excitations together with the $(21)$ component.
Recently similar conclusions were obtained for the many level
pairing model in Ref. \cite{Hir02}.

As a matter of fact it is easy to convince one self that SCRPA is
able to give the exact solution of any two-body problem.
The ground state of a general two particle system can be written as
\bea
|0\ra \sim \left(1+\frac{1}{4}\sum_{p_1h_1p_2h_2}z_{p_1h_1p_2h_2}
a^{\dag}_{p_1}a_{h_1}a^{\dag}_{p_2}a_{h_2}\right)|MF\ra~,
\eea
where the 2p-2h coefficients obey the following symmetry relations
\bea
z_{p_1h_1p_2h_2}=-z_{p_1h_2p_2h_1}=-z_{p_2h_1p_1h_2}=z_{p_2h_2p_1h_1}~.
\eea
Applying the vacuum condition (\ref{annih}) on this state one obtains
\bea
\sum_{ph}\left(-Y^{\nu*}_{ph}+\sum_{p^{\pr}h^{\pr}}
z_{php^{\pr}h^{\pr}}X^{\nu*}_{p^{\pr}h^{\pr}}
\right)a^{\dag}_pa_h|MF\ra =0~,
\eea
or
\bea
Y^{\nu*}_{ph}&=&\sum_{p^{\pr}h^{\pr}}z_{php^{\pr}h^{\pr}}
X^{\nu*}_{p^{\pr}h^{\pr}}~~~{\rm for~all}~\nu~,
\eea
which implies that
\bea
z_{php^{\pr}h^{\pr}}&=&\left[YX^{-1}\right]^*_{php^{\pr}h^{\pr}}~.
\eea
Since we have now expressed the exact ground state in terms of
the SCRPA amplitudes, we also can deduce that the SCRPA excitation
energies are the exact ones because the RPA destruction operator
can be written as $Q_{\nu}=|0\ra\la\nu|$, where $|\nu\ra$ are the
exact state vectors of the two particle excited states.
On the other hand $Q^{\dag}_{\nu}=|\nu\ra\la 0|$ fulfils exactly the
SCRPA equation of motion (\ref{RPAsys}). We therefore see
that SCRPA solves a general two particle problem exactly
as long as one can define a Fermi surface.
This, however, should in principle always be possible in putting
the two particle system in an external box potential
with or without periodic boundary conditions
and solve the SCRPA equations in the limit of an infinitely large box.

\subsection{SCRPA in the spherical region}
\label{subsec:numd}

Let us first discuss the SCRPA results in the spherical region,
i.e. the region where the generalised mean field equation (\ref{HFeq1})
has only the trivial solution $\phi=0$.
In comparison with standard HF this region is strongly extended.
The content of the spherical region depends on the particle number.
We have seen that for $N=2$ the spherical region covers all values
of $\chi$, since we obtained the exact solution.
For $N=10$ or $N=20$ the spherical region is typically extended
by a factor of two. This comes from the selfconsistent coupling
of the quantal fluctuations to the mean field.

Let us now consider the definite example $e_0=0,~e_1=1,~e_2=2$
for $N=20$. In Fig. 3 we show by dashed lines the SCRPA results for the
excitation energies, compared with the exact ones (solid lines)
and to standard RPA (dot-dashes). We see that SCRPA strongly improves
over standard RPA and in fact first and second excited states
are excellently reproduced up to $\chi$-values of about $\chi\approx 1.2$.
The third state has no analogue in standard RPA and must therefore be
attributed to the scattering configuration $(21)$.
The SCRPA solution for the the third eigenvalue definitely seems to
approximate the fifth exact eigenvalue in the range $0\leq\chi\leq1.0$.
The fifth eigenvalue has mainly a 2p-2h structure and it seems
natural that the inclusion of scattering terms allows to reproduce
such states, since e.g. $a^{\dag}_pa_{p^{\pr}}$ in a boson
expansion picture is represented by a 2p-2h configuration, i.e.
\bea
a^{\dag}_pa_{p^{\pr}}\approx \sum_h B^{\dag}_{ph} B_{p^{\pr}h}~,
\eea
where $B_{ph}$ are ideal boson operators (see \cite{Rin80}, Ch. 9).
The fact that SCRPA, via the scattering terms, is able to describe states
of the 2p-2h type is very satisfying and in a way astonishing.
Indeed the single particle energy differences which correspond
to the $a^{\dag}_pa_{p^{\pr}}$ configuration are $e_p-e_{p^{\pr}}$
and in our example this gives $e_2-e_1=1$.
One should have thought that for $\chi\rightarrow 0$ the third
eigenvalue goes to one. In fact it can be shown that for $\chi$
strictly zero, the third eigenvalue {\it is} one.
However, this is a singular point and for any finite value of $\chi$
the third SCRPA root jumps to the values shown in Fig. 3.
A closer inspection reveals that this latter feature is
entirely due to the genuine two-body correlations contained in
SCRPA, because in the r-RPA the third eigenvalue is completely
modified and indeed stays close to $e_2-e_1=1$. We repeated
the calculation for $\Delta\epsilon=e_2-e_1=0.001$.
In Fig. 4 for $N=10$ and Fig. 5 for $N=20$ we see that the scenario
concerning the third eigenvalue stays practically unchanged and,
therefore, this seems to be a quite robust feature. In Fig.
6 we show the r-RPA solution for the $\Delta\epsilon=0.001$ case.
Indeed there, in the spherical region, a very low-lying third
eigenvalue appears which is of the order $\Delta\epsilon$
and thus not distinguishable from the abscissa on the scale of the figure.
We therefore see that the genuine two-body correlations have
dramatic effect in order to restore the situation for the
third eigenvalue.

It is also interesting to investigate the influence of the scattering terms
on the other SCRPA eigenvalues. We therefore performed the calculation
switching off the $(21)$-component in the SCRPA equations for $N=10$ case.
In Table 2 we give the results. It is seen that the influence of
the scattering terms on the first and second SCRPA eigenvalues
is very weak.

\begin{center}
{\bf Table 2}
\vskip5mm
{\it 
The eigenvalues for the two-dimensional $\omega_k^{(2)},~k=1,2$
and three dimensional $\omega_k^{(3)},~k=1,2,3$ versions of the SCRPA
versus the strength $\chi$ (first column) in the spherical region.
The particle number is $N=10$ and $\epsilon_n=n~MeV,~n=0,1,2$.
In the columns 2-4 are given the exact solutions $\omega_k~,k=1,2,5$.
In the last columns are given ground state energies
for two-dimensional and three-dimensional SCRPA and exact values,
respectively.}
\vskip5mm
\begin{tabular}{|c|c|c|c|c|c|c|c|c|c|c|c|}
\hline
 ~~~$\chi$~~~ &
 ~~~$\omega_1$~~~ & ~~~$\omega_2$~~~ & ~~~$\omega_5$~~~ & 
 ~~$\omega_1^{(2)}$~~ & ~~$\omega_2^{(2)}$~~ &
 ~~$\omega_1^{(3)}$~~ & ~~$\omega_2^{(3)}$~~ & ~~$\omega_3^{(3)}$~~ &
 $E^{(2)}_0$ & ~~$E^{(3)}_0$~~ & ~$E^{(exact)}_0$~ \cr
\hline
  0.900 & 0.745 & 1.758 & 2.888 & 0.774 & 1.908 & 0.778 & 1.910 & 3.102 & -0.407 & -0.409 & -0.354 \cr
  0.920 & 0.735 & 1.752 & 2.887 & 0.767 & 1.905 & 0.772 & 1.907 & 3.106 & -0.427 & -0.429 & -0.371 \cr
  0.940 & 0.726 & 1.745 & 2.886 & 0.761 & 1.902 & 0.766 & 1.904 & 3.111 & -0.447 & -0.450 & -0.388  \cr
  0.960 & 0.716 & 1.739 & 2.886 & 0.755 & 1.899 & 0.760 & 1.901 & 3.116 & -0.468 & -0.470 & -0.406 \cr
  0.980 & 0.706 & 1.733 & 2.885 & 0.749 & 1.896 & 0.754 & 1.898 & 3.122 & -0.490 & -0.492 & -0.423 \cr
  1.000 & 0.696 & 1.727 & 2.885 & 0.744 & 1.893 & 0.749 & 1.895 & 3.128 & -0.512 & -0.514 & -0.442 \cr
  1.020 & 0.686 & 1.722 & 2.885 & 0.738 & 1.889 & 0.744 & 1.891 & 3.135 & -0.534 & -0.536 & -0.461 \cr
  1.040 & 0.677 & 1.717 & 2.886 & 0.733 & 1.886 & 0.739 & 1.888 & 3.143 & -0.557 & -0.559 & -0.480 \cr
  1.060 & 0.667 & 1.712 & 2.886 & 0.728 & 1.883 & 0.734 & 1.885 & 3.151 & -0.580 & -0.582 & -0.500 \cr
  1.080 & 0.657 & 1.707 & 2.887 & 0.723 & 1.880 & 0.729 & 1.882 & 3.160 & -0.604 & -0.606 & -0.520 \cr
  1.100 & 0.647 & 1.703 & 2.889 & 0.719 & 1.877 & 0.725 & 1.879 & 3.169 & -0.628 & -0.631 & -0.540 \cr
\hline
\end{tabular}
\end{center}
\vskip5mm

In Figs. 7 and 8 we show the SCRPA groundstate energies for $N=10$
and $N=20$, respectively as a function of the angle $\phi$
representing the mean field transformation.
We take $\Delta\epsilon=0.001$. We see that for a large plateau
of $\chi$-values the minimum stays at $\chi=0$. This region is
more or less enhanced by a factor of two, compared with standard
HF theory, where the $\phi=0$ minimum ceases to exist at $\chi=1$.
In SCRPA, after the $\phi=0$ solution is finished, the system
jumps in a discontinuous manner to the "new" deformed solution.
This discontinuity is clearly seen in Figs. 9 ($N=10$) and 10
($N=20$) where we show the ground state energy $E_0^{(SCRPA)}$ (dashed line)
and compare it with the exact values (solid line) and standard RPA
(dot-dashed line). The full 3x3 SCRPA with the scattering terms
was used. We see a strong improvement of SCRPA versus standard RPA
in the transition region where standard HF switches from the
spherical to deformed regime.
Beyond this transition point SCRPA stays spherical and slightly
overbinds whereas standard RPA goes deformed and slightly
underbinds. From the point where also SCRPA jumps to the
deformed solution, standard and SCRPA give almost identical
correlation energies.

\subsection{SCRPA in the deformed region}
\label{subsec:nume}

In the deformed region a particular situation arises in our model
for $\Delta\epsilon=0$, since, as already mentioned, a spontaneously
broken symmetry occurs in this case. Here the standard HF-RPA
exhibits its real strength because, as shown in (\ref{RPAfre1}),
a zero mode appears which signifies that the broken symmetry is
restored, i.e. the conservation laws are fulfilled 
\cite{Bla86,Rin80}.
We will now demonstrate that this property is conserved in SCRPA
(with scattering terms). In this context we mention that the
symmetry operator
\bea
\label{L0}
\hat{L}_0=i(K_{21}-K_{12})=i\left[A_{20}-A_{02}){\rm sin}\phi+
(A_{21}-A_{12}){\rm cos}\phi\right]~,
\eea
which commutes with the Hamiltonian, can be identified with the
z-component of the rotation operator. This operator contains
scattering terms which are, however, also present in our RPA-ansatz
and therefore we can expect that (\ref{L0}) is a particular
RPA operator with a zero eigenvalue solution in SCRPA, very much in the
same way as this is the case in standard RPA.
The numerical verification of this desirable quality of SCRPA
must however be undertaken with care. Indeed, a zero mode contains
diverging amplitudes which, injected into the SCRPA matrix,
may not lead to selfconsistency.
The way to overcome this difficulty is to start the calculation with
a finite small value of $\Delta\epsilon$, i.e. with a slight explicit
symmetry breaking, and then to diminish its value step by step.
We, in this way, could verify with very high accuracy that the zero
eigenvalue occurs in the deformed region for all values of the
interaction strength $\chi$. This is shown in Fig. 4 
by a solid line for $N=10$ and Fig. 5 for $N=20$.
Here we considered the value $\Delta\epsilon=0.001$, but we were
able to reach the value $\Delta\epsilon=10^{-6}$.

In spite of this clear appearance of the Goldstone mode,
which only shows up when scattering terms are included
and the generalised mean field equation (\ref{HFeq1}) is solved
selfconsistently with the SCRPA equations, the spectrum in the
deformed region needs some discussion and clarification.
First of all one notices a discontinuity when passing from
the spherical to the deformed region. This feature of
deformed SCRPA has been already noticed earlier with other models:
the two level Lipkin model \cite{Duk90} and the two level
pairing model \cite{Rab02}.
It is reminiscent of a first order phase transition which,
however, is absent in the exact solution.

Furthermore the second SCRPA eigenvalue seems too low compared with
the exact solution and also, astonishingly, with respect to the standard RPA
solution, as seen in Figs. 1,2. In fact standard RPA works surprisingly
well in the deformed region, a fact which seems to be a constancy
in all the models we have investigated so far \cite{Duk90,Rab02}.

Third, in the deformed region it does not seem easy to identify
the third SCRPA eigenvalue with any of the exact solutions.
Therefore, in spite of satisfying the Goldstone theorem,
the interpretation of SCRPA seems less clear in the deformed region
than it appears in the spherical one. However, the situation may be less
unsatisfactory than it appears at first sight.

As usual with a continuously broken symmetry also in the present
model a clear rotational band structure is revealed, as can be
seen by inspecting Figs. 1,2. 
The exact solution, found by a diagonalisation procedure,
has a definite angular momentum projection $L_0$.
Moreover the expectation value of the $L_0^2$ operator
has integer values, namely
\bea
\label{band}
\sqrt{\langle L_0^2\rangle}=J=0,1,2,...
\eea
The first "rotational band" $J=0,1,2,...$ is built on top
of the RPA excitation with a vanishing energy (Goldstone mode).
The second "band-head" corresponds to the second RPA mode (dashed curve)
and the corresponding members of this "excited band" have a
different slope with respect to the "ground band".
On the same figures one can also see a third "rotational band".

As customary in RPA theory one also can evaluate the mass
parameter of the rotational band within SCRPA.
By a straightforward generalisation we obtain for the
moment of inertia (see e.g. Ref. \cite{Bla86,Rin80})
\bea
\label{mass}
M=2L_0^*\left({\cal A}-{\cal B}\right)^{-1}L_0~,
\eea
where ${\cal A,B}$ are the SCRPA matrices and
$L_0$ denotes the part with $m>i$ components of the momentum
operator (\ref{L0}),
which should be written in terms of normalised generators
$\delta Q^{\dag}$ defined by Eq. (\ref{gen}), i.e.
\bea
\label{L01}
L_0=i\left(0,~N_{20}^{1/2}sin\phi,~N_{21}^{1/2}cos\phi
\right)~.
\eea
For the standard RPA case, by using the matrix elements given by
(\ref{RPA}), and $N_{20}=N$, one obtains an analytical solution, namely
\bea
\label{mass0}
M=\frac{N(\chi-1)}{\epsilon\chi(\chi+1)}~.
\eea
In Fig. 11 we show by a solid line the above moment of inertia calculated
with RPA for $N=20$ and in Fig. 12 with SCRPA.
The comparison with the "exact" mass parameter fitted from the
exact spectrum is also given by dashed curves. 
From the dot-dashed lines, giving their ratio, we see that
the SCRPA mass is slightly closer to the "exact" value
than the corresponding RPA value. This is also manifest when
comparing the rotational energies, as in Fig. 13.
Here we plotted by dashes the SCRPA spectrum and by solid lines
the "exact" values.

One therefore sees that one can recover rather clearly the precise
position of the lowest energy value and therefore the discontinuity,
at least of the first excited state, in crossing the phase boundary,
is much less pronounced, than the one which is seen in Figs. 4, 5.
One may think that the same upward shift should also be observed
for the second SCRPA eigenvalue (the band head of the second
rotational band) bringing it in a close agreement with the
exact solution.
It then appears that one can use SCRPA with scattering terms
in the same way as standard RPA with all usual properties conserved.
Let us mention that even in r-RPA the Goldstone mode appears when
the scattering terms are considered. This is shown in Fig. 6.
Note, however, that in the spherical region the third r-RPA eigenvalue
also comes almost zero value. The reason for this was discussed above.

\subsection{The occupation numbers}
\label{subsec:numf}

The reproduction of energies may not be the most sensitive
indicator for the performance of a theory.
Usually wavefunctions are much more demanding in this respect.
We therefore calculated for $\Delta\epsilon=0.001$ the occupation
numbers which reflect more directly the quality of the wave function,
since they are not obtained from a minimisation procedure, as the
energies.

In Figs. 14, 15 we show the SCRPA results for 
$\la 0|\rho_0|0\ra,~\la 0|\rho_1|0\ra$ and $\la 0|\rho_2|0\ra$
according to Eq. (\ref{denlin}).
We see that in the spherical region the SCRPA densities (solid lines)
are rather close to the exact values (dashes) for $\chi$-values
up to slightly above one. This is, of course, much better than standard
RPA (not shown) which breaks down, as the energies, well
before $\chi=1$, but the region $1\leq\chi\leq 2$ remains not well
reproduced. On the contrary in the deformed region the SCRPA
results are excellent. It is somewhat unexpected to see that the
occupation numbers of the two upper but degenerated levels are
not equal but their averaged sum very well reproduce
the exact occupancy of the two upper levels (dotted lines in
Figs. 14, 15).
Therefore the SCRPA wavefunction performs well in the deformed region.


\section{Conclusions}
\label{sec:numer}
\setcounter{equation}{0}
\renewcommand{\theequation}{4.\arabic{equation}}

In this paper we studied various aspects of the SCRPA in the
three level Lipkin model. With respect to the two level
Lipkin model, the three level version has the advantage of allowing
for a continuously broken symmetry on the mean field level
with the appearance of a Goldstone mode.
One major objective of the present work therefore was to investigate
whether in SCRPA the very desirable properties of standard RPA
concerning exact restoration of symmetry and fullfillment of
conservation laws can also be maintained.
We found out that this is indeed the case under the condition that the
symmetry operator is fully included in the RPA operator as a particular
solution. In our model this implies that the RPA operator contains
in addition to the usual ph and hp components 
$a^{\dag}_pa_h$ and $a^{\dag}_ha_p$ also the so-called anomalous or
scattering terms $a^{\dag}_pa_{p^{\pr}}$ and $a^{\dag}_ha_{h^{\pr}}$.

With a correlated ground state the inclusion of such terms
is perfectly possible since the occupation numbers are now
different from one and/or zero and mathematically the situation
becomes similar to the finite temperature RPA case, where
also $pp^{\pr}$ and $hh^{\pr}$ components are allowed.
However, it was found in this work that the inclusion of
such scattering terms is not without problem.
For example $pp^{\pr}~(hh^{\pr})$ configurations
can imply very small energy differences
$\epsilon_p-\epsilon_{p^{\pr}}\approx 0$ and therefore for small
interaction very low-lying roots may be found in the spectrum
which are absent in the exact solution.
In our model it turned out that this is the case only in the approximate,
so-called renormalised version of SCRPA.
On the contrary, in the full SCRPA the eigenvalue which is linked
to the inclusion of scattering terms corresponds to a 2p-2h
state of the exact spectrum. This is true in the "spherical" region
whereas in the symmetry broken phase the third eigenvalue
does not seem to have any precise correspondent in the exact solution.
In the symmetry unbroken phase the elimination of the
anomalous terms has an insignificant influence on the reminder
of the spectrum. In the symmetry broken phase this inclusion
is crucial to produce the Goldstone mode at zero energy,
otherwise this state is pushed up to quite high energy with
no equivalent in the exact spectrum.
We also calculated the moment of inertia of the rotational band
which works in SCRPA (with scattering terms) in a way very analogous
to standard RPA. Very good agreement with the exact solution is
found. Therefore the present formulation of SCRPA allows to maintain
all the formal and desirable properties of standard RPA.
There is only one unpleasant feature which is that SCRPA solutions
do not join smoothly when passing from the symmetry unbroken to the
symmetry broken phase. Whether this can be improved upon remains to
be seen in future work. A further very attractive feature of SCRPA
is the fact that it solves any two body problem {\it exactly}.
This is contrary to the usual where many body approximations
deteriorate when decreasing the number of particles.

Quite remarkably the SCRPA solution exists in the
symmetry unbroken single particle basis well beyond the phase transition
point given by the standard RPA and the excited state is well
approximated in all this region.
For $N=10$ and $N=20$ at a certain strength
$\ov{\chi}_{crt}\approx 2\chi_{crt}$,
where $\chi_{crt}$ is the value where standard HF becomes unstable,
SCRPA also jumps to the deformed solution.
A similar behaviour of SCRPA has already been found for other
models.

Though the appearance of the Goldstone mode is a very desirable
feature, since this signals that conservation laws and sum rules
are fulfilled, it turned out that in the deformed region,
contrary to the "spherical" one, the SCRPA only very slightly
improves over standard RPA and both, SCRPA and standard RPA
are quite close to the exact solution.
Again this is a feature which also has been found in other models.
It seems that the common wisdom that "deformation"
sums already a lot of additional correlations in standard RPA
is born out here. 

We therefore can summarize the studies of SCRPA in various models
in the following way: in the symmetry unbroken regime SCRPA
works excellently, yielding simultaneously very good ground state
and excitation energies. One can work in the spherical basis
beyond the usual transition point given by the instability
of the standard HF solution. This "beyond" depends on the
number of particles and the model.
SCRPA solves a general two particle system {\it exactly}.

In the deformed region SCRPA does not seem to improve very much
over standard "deformed" RPA. However, all formal properties
of standard RPA are maintained with SCRPA when the scattering terms
are included. For example for nuclei SCRPA will give a Goldstone mode
restoring translational invariance.
The transition from "spherical" to "deformed"
is discontinuous, which is an unpleasant feature.
It is not evident that the inclusion of scattering components
in the {\it spherical region} improves the results.
It also should be mentioned that the appearance of the Goldstone
mode in SCRPA is only guaranteed when the symmetry operator
does not contain any diagonal components as $a^{\dag}_ia_i$,
since those terms cannot be included in the RPA operator.
Such a situation occurs for example in the case of
particle number violation \cite{Rin80}.

\vskip1cm

{\rm {\bf Acknowledgements} }

{\rm One of us (D.S.D.) is grateful for the financial support given by
IPN Orsay, were part the work was performed.
J.D. gratefully acknowledges partial support by the Spanish DGI
under Grant BFM2003-05316-C02-02.
Discussions with M. Sambataro, M.Grasso (Catania)
and N.V. Giai (Orsay) are gratefully acknowledged.}

\newpage
\centerline{\bf Appendix A}
\centerline{\bf SCRPA in two steps}
\setcounter{equation}{0}
\renewcommand{\theequation}{A.\arabic{equation}}
\vskip5mm

The SCRPA system of equations (\ref{RPAsys}) 
and the normalisation condition for amplitudes
can be written in a matrix form respectively  as follows
\bea
\label{RPAsys1}
{\cal S X} &=& \ov{\cal X} \Omega~,
\nn
{\cal X}^{\dag}\ov{\cal X} &=& {\cal I}~.
\eea\
Here we introduced some short-hand notations
\bea
{\cal S} = 
\left(\begin{matrix}
a & b \cr b^* & a^* \cr
\end{matrix}\right)~,~~~
{\cal X} = 
\left(\begin{matrix}
x & y^* \cr y & x^* \cr
\end{matrix}\right)~,~~~
\ov{\cal X}={\cal N X}~,
\nn
{\cal N} = 
\left(\begin{matrix}
N & 0 \cr 0 & -N \cr
\end{matrix}\right)~,~~~
{\cal I} = 
\left(\begin{matrix}
I & 0 \cr 0 & -I \cr
\end{matrix}\right)~,~~~
\Omega = 
\left(\begin{matrix}
\omega & 0 \cr 0 & -\omega \cr
\end{matrix}\right)~.
\eea
Here the matrices $a,~b$ are defined by Eq. (\ref{ARPA}).
We used the amplitudes ${\cal X}$ relative to the initial generators
$A_{mi}$ defined by Eq. (\ref{redamp}) and also the SCRPA matrices defined
by Eq. (\ref{ARPA}).
We also introduced similar to (\ref{redamp2}) amplitudes $\ov{\cal X}$.
$N$ is the diagonal metric matrix $N_{k}\delta_{kl}$,
where we used for the basis indices the notation $(mi)\rightarrow k$
\bea
\label{notation}
(10)\rightarrow 1,~~~(20)\rightarrow 2,~~~(21)\rightarrow 3~.
\eea
SCRPA equations are written as
\bea
{\cal N} \ov{\cal S} \ov{\cal X} &=& \ov{\cal X} \Omega~,
\eea
in terms of the SCRPA matrix relative to $\ov{\cal X}$ amplitudes
\bea
\ov{\cal S}={\cal N}^{-1}{\cal S}{\cal N}^{-1}~.
\eea
From the normalisation condition one derives the inverse
amplitude matrix
\bea
{\cal X}^{-1}={\cal I}\ov{\cal X}^{\dag}~,
\eea
giving the following resolution of the unity
\bea
\label{unity}
{\cal X I}\ov{\cal X}^{\dag} = {\bf I} \equiv
\left(\begin{matrix}I & 0 \cr 0 & I \cr\end{matrix}\right)~.
\eea
Let us split the SCRPA matrix into two parts: the r-RPA matrix
plus a fluctuation
\bea
{\cal S} = {\cal S}_0 + \delta {\cal S}~,
\eea
where $\delta {\cal S} = {\cal S} -{\cal S}_0$.
The r-RPA system of equations and the normalisation condition
are given by
\bea
\label{RPAsys0}
{\cal S}_0 {\cal X}_0 &=& \ov{\cal X}_0 \Omega_0~,
\nn
{\cal X}_0^{\dag}\ov{\cal X}_0 &=& {\cal I}~,
\nn
\ov{\cal X}_0 &=& {\cal N X}_0~.
\eea
We should stress on the important fact that our calculation is
selfconsistent because we use the SCRPA set of amplitudes in deriving
the metric matrix ${\cal N}$ in the r-RPA system of equations.
We insert the unity operator written in terms of r-RPA amplitudes
\bea
\label{unity0}
{\cal X}_0 {\cal I}\ov{\cal X}^{\dag}_0 = {\bf I}~,
\eea
in front of ${\cal X}$ amplitudes of the left hand side Eq.
(\ref{RPAsys1}). By using r-RPA equations (\ref{RPAsys0}) one gets
\bea
\left( \ov{\cal X}_0 \Omega_0 +\delta {\cal S X}_0 \right)
{\cal I X}_0^{\dag}\ov{\cal X} = \ov{\cal X}\Omega~.
\eea
We multiply this relation to the left with 
$\ov{\cal X}_0^{-1}={\cal I X}^{\dag}_0$.
One finally obtains a standard system of RPA equations
\bea
\left( \Omega_0 {\cal I} + \delta {\cal S}^{\pr} \right)
{\cal X}^{\pr} ={\cal I X}^{\pr}\Omega~,
\eea
with the normalisation condition for new amplitudes
\bea
\label{norme2}
{\cal X}^{\pr\dag} {\cal I X}^{\pr} &=& {\cal I}~,
\eea
where
\bea
{\cal X}^{\pr} &=& {\cal X}_0^{\dag}\ov{\cal X}~,
\nn
\delta {\cal S}^{\pr} &=&
{\cal I X}_0^{\dag} \delta {\cal S} {\cal X}_0 {\cal I}~.
\eea
In particular one has
\bea
\delta a^{\pr}&=&
 x_0^{\dag}\delta a x_0
+x_0^{\dag}\delta b y_0
+y_0^{\dag}\delta b x_0
+y_0^{\dag}\delta a y_0~,
\nn
\delta b^{\pr}&=&
-x_0^{\dag}\delta a y_0
-x_0^{\dag}\delta b x_0
-y_0^{\dag}\delta b y_0
-y_0^{\dag}\delta a x_0~.
\eea
This is an RPA system of equations with r-RPA energies on the diagonal.
It corresponds to a new representation of the SCRPA phonon
\bea
Q^{\dag}_{\nu}=\sum_{\alpha}\left(
x^{\pr\nu}_{\alpha}Q^{\dag}_{0\alpha}-
y^{\pr\nu}_{\alpha}Q_{0\alpha}\right)~,
\eea
in terms of r-RPA phonons
\bea
Q^{\dag}_{0\alpha}=\sum_{m>i}\left(
x^{\alpha}_{0k}A_{mi}-
y^{\alpha}_{0k}A_{im}\right)~.
\eea

The formulation of SCRPA, where the division by the norm matrix
${\cal N}$ has disappeared, has the important formal aspect
that one sees that the division by eventual very small numbers 
or even zero, contained in ${\cal N}$, in reality does not exists.
This is also born out by our three component SCRPA solution
treated in the main text.

\newpage
\centerline{\bf Appendix B}
\centerline{\bf Operator expansion method for densities}
\setcounter{equation}{0}
\renewcommand{\theequation}{B.\arabic{equation}}
\vskip5mm

In order to find the coefficients of the expansion (\ref{expans})
it is possible to use the resolution of unity method \cite{Sch71,Feu00}.
We proceed in a more direct way by computing the matrix elements
with respect the complete basis (\ref{Abasis}).
One obtains a lower triangular system of equations
\be
\label{syst}
\sum_{n_1\leq m_1}\sum_{n_2\leq m_2}
{\cal M}_{m_1m_2}^{n_1n_2}c_{n_1n_2}(k_0k_1k_2)=
{\cal L}_{m_1m_2}(k_0k_1k_2)~,
\ee
where
\bea
\label{coef}
{\cal M}_{m_1m_2}^{n_1n_2}&\equiv&{\cal N}^{-1}_{m_1m_2}
\langle 0|A^{m_2}_{02}A^{m_1}_{01}A^{n_1}_{10}A^{n_2}_{20}
A^{n_2}_{02}A^{n_1}_{01}A^{m_1}_{10}A^{m_2}_{20}|0\rangle~,
\nn
{\cal L}_{m_1m_2}(k_0k_1k_2)&\equiv&{\cal N}^{-1}_{m_1m_2}
\langle 0|A^{m_2}_{02}A^{m_1}_{01}
\hat{\rho}_0^{k_0}\hat{\rho}_1^{k_1}\hat{\rho}_2^{k_2}
A^{m_1}_{10}A^{m_2}_{20}|0\rangle~.
\eea
The solution of this system is given by the following recurrent relation
\be
\label{sol}
c_{m_1m_2}(k_0k_1k_2)=\frac{1}{{\cal M}_{m_1m_2}^{m_1m_2}}
\left[{\cal L}_{m_1m_2}(k_0k_1k_2)-\sum_{n_1n_2<m_1m_2}
{\cal M}_{m_1m_2}^{n_1n_2}c_{n_1n_2}(k_0k_1k_2)
\right]~.
\ee
Let us consider the HF ansatz of the vacuum, defined by
\be
a^{\dag}_{i\mu}a_{j\nu}|HF\rangle=
\delta_{ij}\delta_{\mu\nu}\delta_{i0}|HF\rangle~.
\ee
The coefficients of the system (\ref{syst}) are given in a straightforward
way
\bea
{\cal L}_{m_1m_2}(k_0k_1k_2)
&=&(N-m_1-m_2)^{k_0}m_1^{k_1}m_2^{k_2}
\nn
{\cal M}_{m_1m_2}^{n_1n_2}
&=&{\cal N}_{m_1m_2}
\frac{(N-m_1-m_2+n_1+n_2)!}{N!(m_1-n_1)!(m_2-n_2)!}
\nn
{\cal N}_{m_1m_2}
&=&\frac{N!m_1!m_2!}{(N-m_1-m_2)!}
\nn
c_{00}(k_0k_1k_2)&=&N^{k_0}\delta_{k_10}\delta_{k_20}~.
\label{HFcoef}
\eea
We give the solutions of this system for $n_1+n_2\leq 2$
\bea
\hat{\rho}_0^k&\approx&N^k+\frac{(N-1)^k-N^k}{N}(A_{10}A_{01}+A_{20}A_{02})
\nn
&+&\frac{1}{2N(N-1)}\left[(N-2)^k+N^{k-1}(N-2)-\frac{2(N-1)^{k+1}}{N}\right]
\nn
&\times&(A_{10}^2A_{01}^2+A_{20}^2A_{02}^2
+2A_{10}A_{20}A_{02}A_{01})
\nn
\hat{\rho}_m^k&\approx&\frac{1}{N}A_{m0}A_{0m}
+\frac{1}{N(N-1)}\left[(2^{k-1}-1+\frac{1}{N})
A_{m0}^2A_{0m}^2+\frac{1}{N}A_{10}A_{20}A_{02}A_{01}\right]
\nn
\hat{\rho}_0\hat{\rho}_m&\approx&\frac{N-1}{N}A_{m0}A_{0m}
-\frac{1}{N^2(N-1)}(A_{m0}^2A_{0m}^2+A_{10}A_{20}A_{02}A_{01})
\nn
\hat{\rho}_1\hat{\rho}_2&\approx&\frac{1}{N(N-1)}A_{10}A_{20}A_{02}A_{01}~,
\label{expans0}
\eea
where $m=1,2$.
These expansions are very convergent.
Indeed, this can be seen in Table 3, where there are given the expansion
coefficients $c_{n_1n_2}(k_0k_1k_2)$
with $n_1+n_2\leq 2$ for $N=10$.

In order to find the normalisation factors $N_{i0}$ using the
expansion (\ref{expans}) up to $n_1+n_2\leq 2$ we need the expectation
values for the following products of operators
\bea
\langle 0|A_{m0}A_{0m}|0\rangle&=&
N_{m0}\sum_{\nu}Y^{\nu}_{m0}Y^{\nu}_{m0}
\nn
&\equiv& N_{m0}y_{mm}~,
\nn
\langle 0|A_{m0}A_{n0}A_{0n}A_{0m}|0\rangle&=&N_{m0}N_{n0}\sum_{\nu\mu}
[  Y^{\nu}_{m0}Y^{\mu}_{m0}X^{\nu}_{n0}X^{\mu}_{n0}
\nn
&+&Y^{\nu}_{m0}Y^{\nu}_{m0}Y^{\mu}_{n0}Y^{\mu}_{n0}
 + Y^{\nu}_{m0}Y^{\mu}_{m0}Y^{\nu}_{n0}Y^{\mu}_{n0}]
\nn
&\equiv&N_{m0}N_{m0}z_{mn}~.
\eea
For $n_1+n_2\leq 2$ one obtains the following system
in terms of the expansion coefficients in Eq. (\ref{expans})
\bea
\label{sysN}
N_{10}&=&
c_{00}(100)-c_{00}(010)+
[c_{10}(100)-c_{10}(010)]N_{10}y_{11}+
[c_{01}(100)-c_{01}(010)]N_{20}y_{22}
\nn&+&
[c_{20}(100)-c_{20}(010)]N_{10}^2z_{11}+
[c_{02}(100)-c_{02}(010)]N_{20}^2z_{22}
\nn&+&
[c_{11}(100)-c_{11}(010)]N_{10}N_{20}z_{12}~,
\nn
N_{20}&=&
c_{00}(100)-c_{00}(001)+
[c_{10}(100)-c_{10}(001)]N_{10}y_{11}+
[c_{01}(100)-c_{01}(001)]N_{20}y_{22}
\nn&+&
[c_{20}(100)-c_{20}(001)]N_{10}^2z_{11}+
[c_{02}(100)-c_{02}(001)]N_{20}^2z_{22}
\nn&+&
[c_{11}(100)-c_{11}(001)]N_{10}N_{20}z_{12}~.
\eea


\begin{center}
{\bf Table 3}
\vskip5mm
{\it The expansion coefficients $c_{n_1n_2}(k_0k_1k_2)$
in Eq. (\ref{expans}) with $n_1+n_2\leq 2$ for $N=10$.}
\vskip5mm
\begin{tabular}{|r|r|r|r|r|r|r|}
\hline
 $n_1,n_2$&  0,0~~&  0,1~~&  1,0~~&  0,2~~&  1,1~~&  2,0~~\cr
$k_1,k_2,k_3$~&   &       &       &       &       &       \cr
\hline
 1,0,0~& 10.0000&-0.1000&-0.1000&-0.0011&-0.0022&-0.0011\cr
 0,1,0~&  0.0000& 0.0000& 0.1000& 0.0000& 0.0011& 0.0011\cr
 0,0,1~&  0.0000& 0.1000& 0.0000& 0.0011& 0.0011& 0.0000\cr
 2,0,0~&100.0000&-1.9000&-1.9000&-0.0100&-0.0200&-0.0100\cr
 0,2,0~&  0.0000& 0.0000& 0.1000& 0.0000& 0.0011& 0.0122\cr
 0,0,2~&  0.0000& 0.1000& 0.0000& 0.0122& 0.0011& 0.0000\cr
 1,1,0~&  0.0000& 0.0000& 0.9000& 0.0000&-0.0011&-0.0011\cr
 0,1,1~&  0.0000& 0.0000& 0.0000& 0.0000& 0.0111& 0.0000\cr
 1,0,1~&  0.0000& 0.9000& 0.0000&-0.0011&-0.0011& 0.0000\cr
\hline
\end{tabular}
\end{center}



\newpage

\begin{figure}[p]
\begin{center}
\includegraphics[]{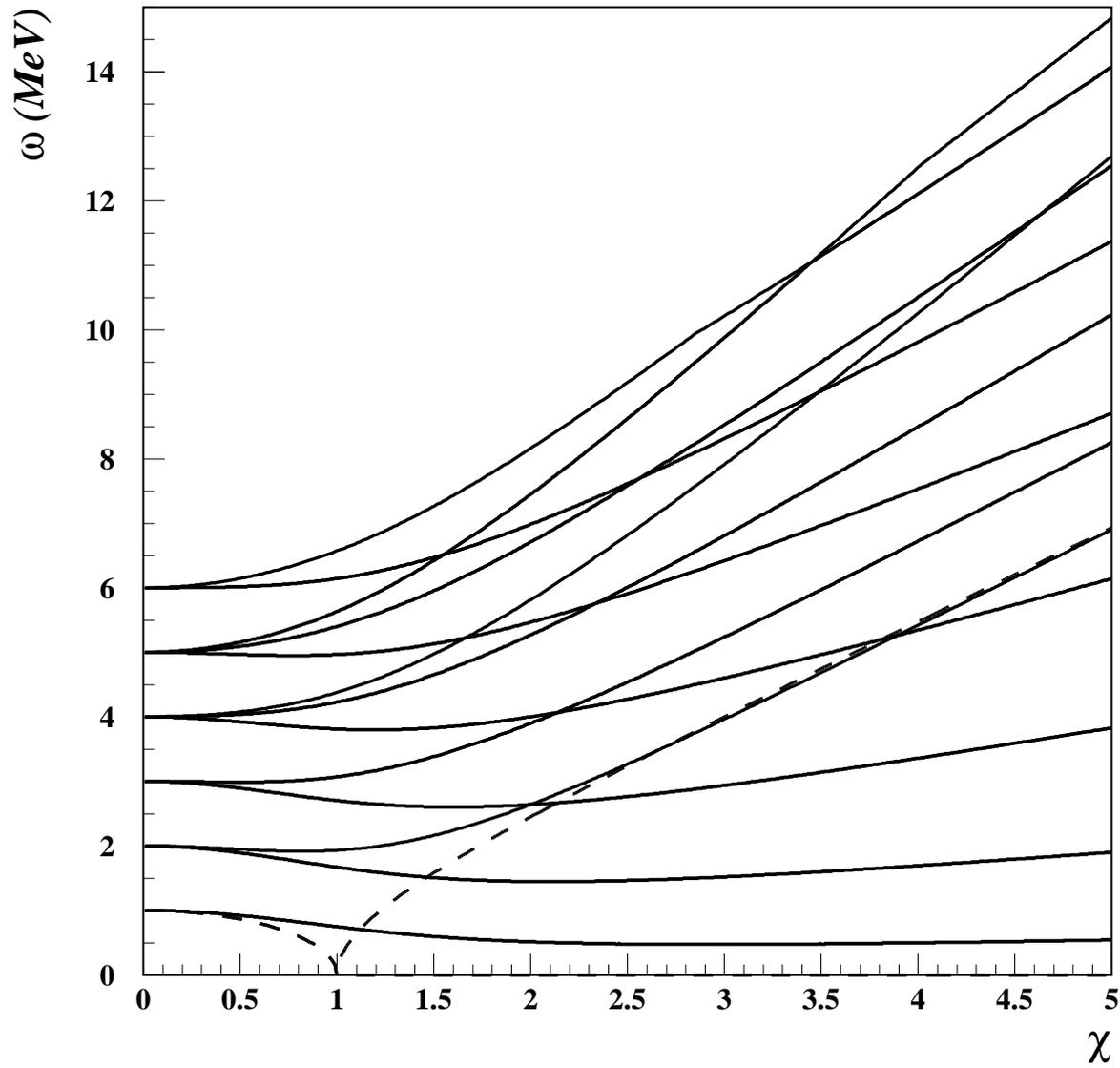}
\vspace{1cm} \caption
{
Exact excitation energies versus the strength parameter $\chi$,
for $N=10$, $\epsilon_1=0, \epsilon_2=\epsilon_3=1 MeV$.
By dashes are given standard RPA values.
Notice the appearance of a zero mode solution (Goldstone mode)
beyond $\chi$=1.
}
\label{fig01}
\end{center}
\end{figure}

\begin{figure}[p]
\begin{center}
\includegraphics[]{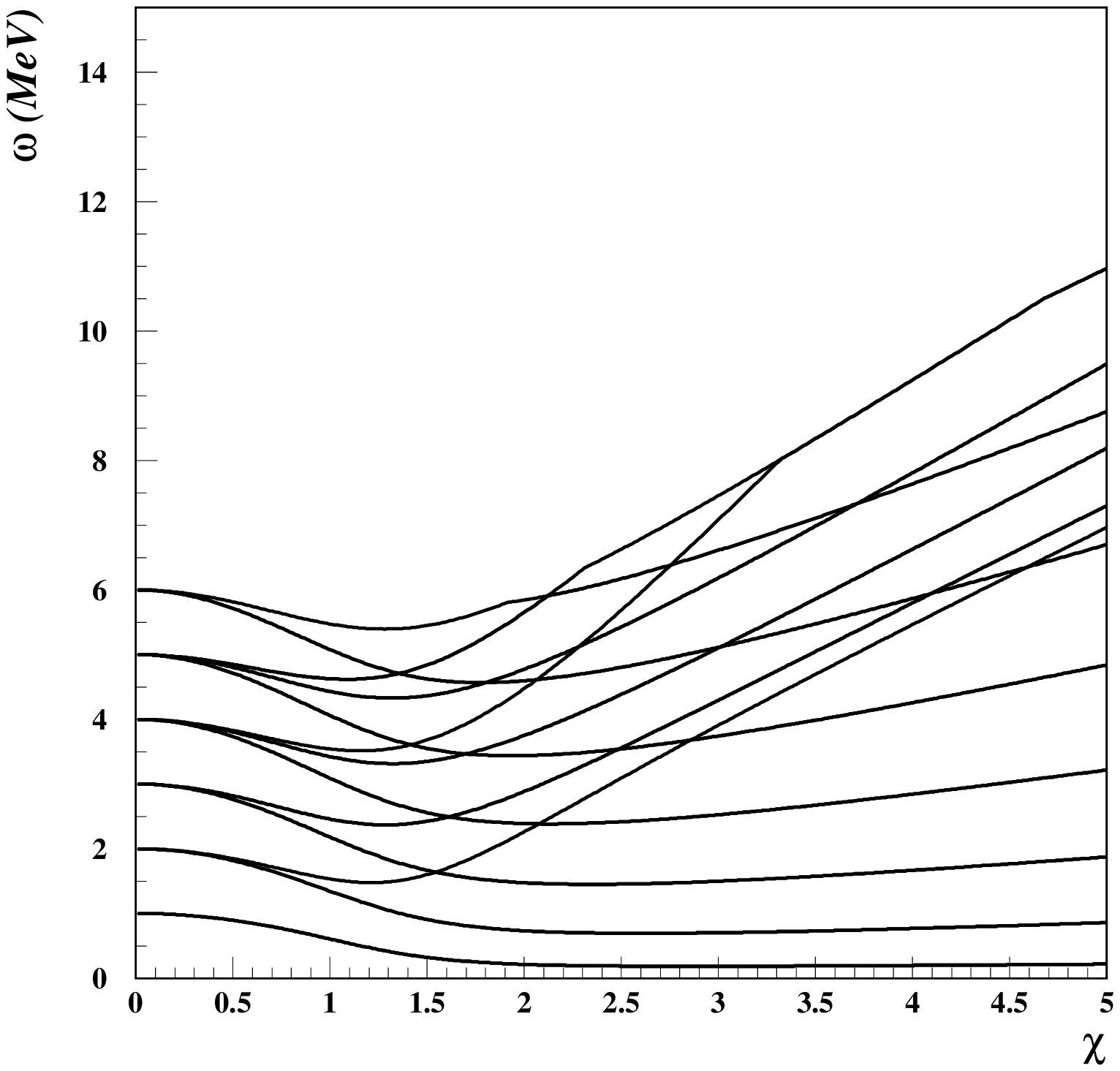}
\vspace{1cm} \caption
{
The same as in Fig. 1, but for $N=20$.
}
\label{fig02}
\end{center}
\end{figure}

\begin{figure}[p]
\begin{center}
\includegraphics[]{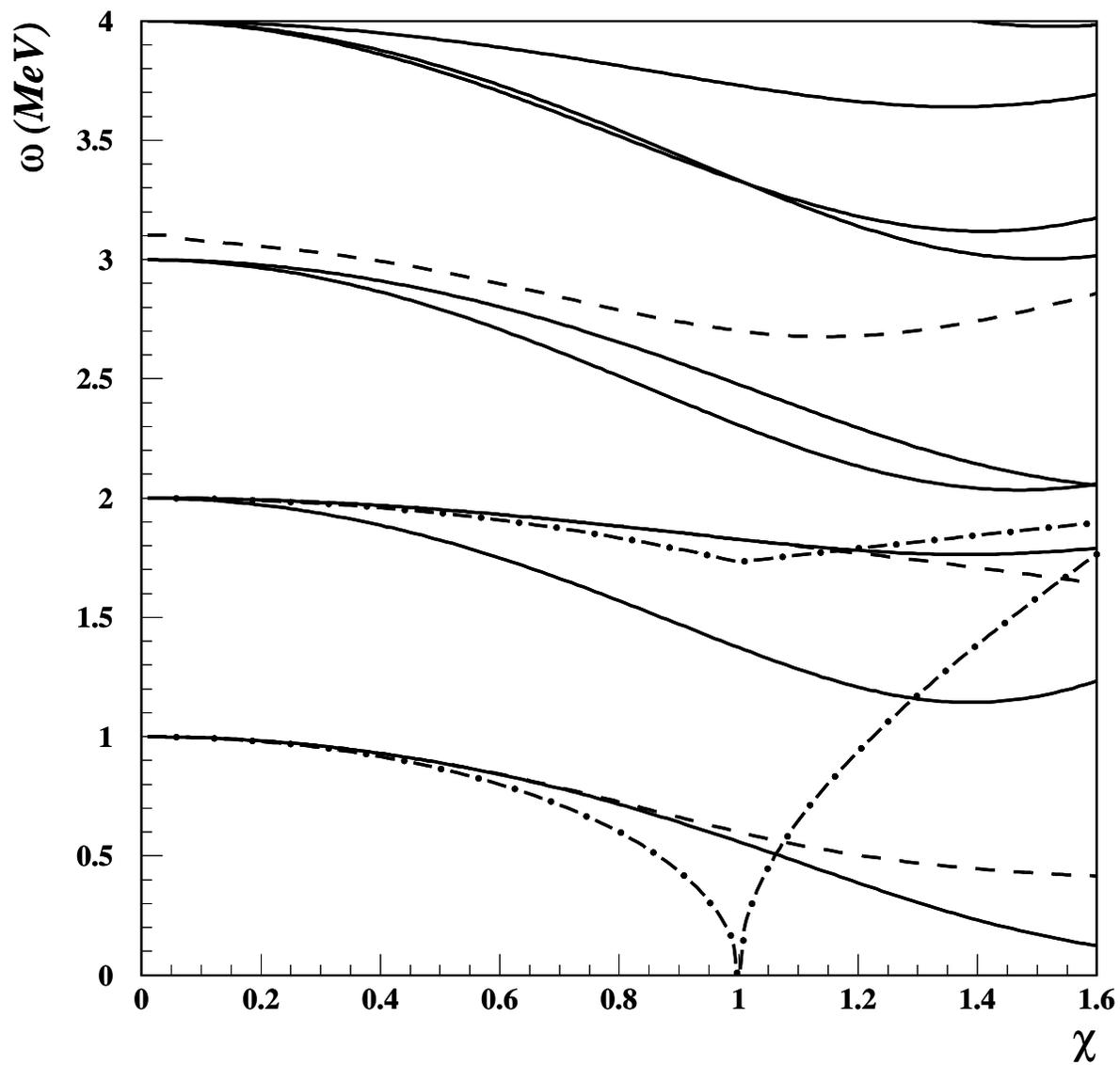}
\vspace{1cm} \caption
{
SCRPA excitation energies versus the strength parameter $\chi$,
for $N=20$ and $e_0=0,~e_1=1,~e_2=2$ (dashed lines).
By solid lines are given the lowest exact eigenvalues and
by dot-dashes the standard RPA energies.
}
\label{fig03}
\end{center}
\end{figure}

\begin{figure}[p]
\begin{center}
\includegraphics[]{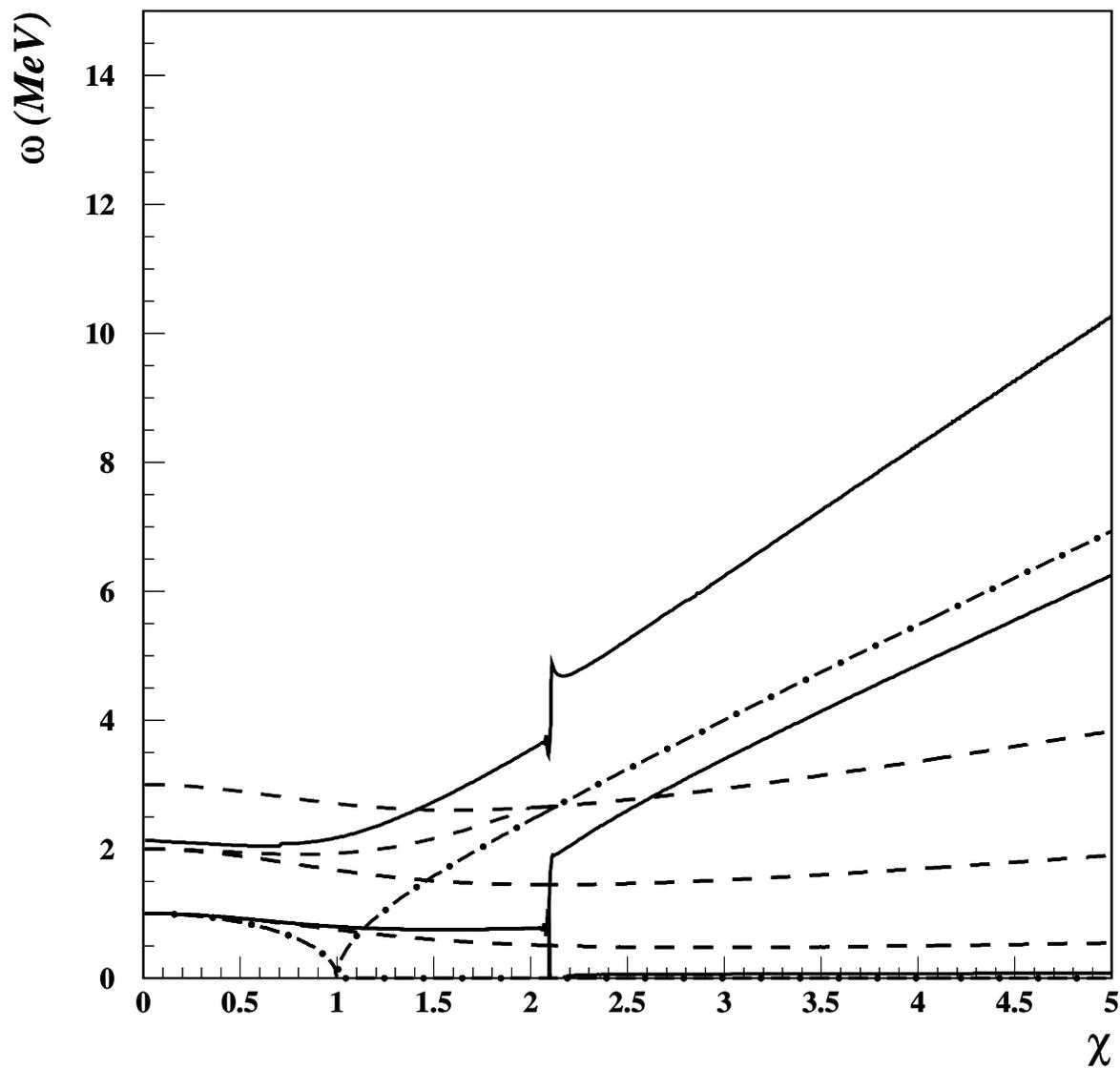}
\vspace{1cm} \caption
{
SCRPA excitation energies versus the strength parameter $\chi$,
for $N=10$, $\Delta\epsilon=0.001~MeV$ (full line).
By dashes are given the lowest exact eigenvalues and
by dot-dashes the standard RPA energies.
After the phase transition point $\chi=1$ (standard RPA) and $\chi\approx 2.1$
(SCRPA) a Goldstone mode at zero energy appears.
}
\label{fig04}
\end{center}
\end{figure}

\begin{figure}[p]
\begin{center}
\includegraphics[]{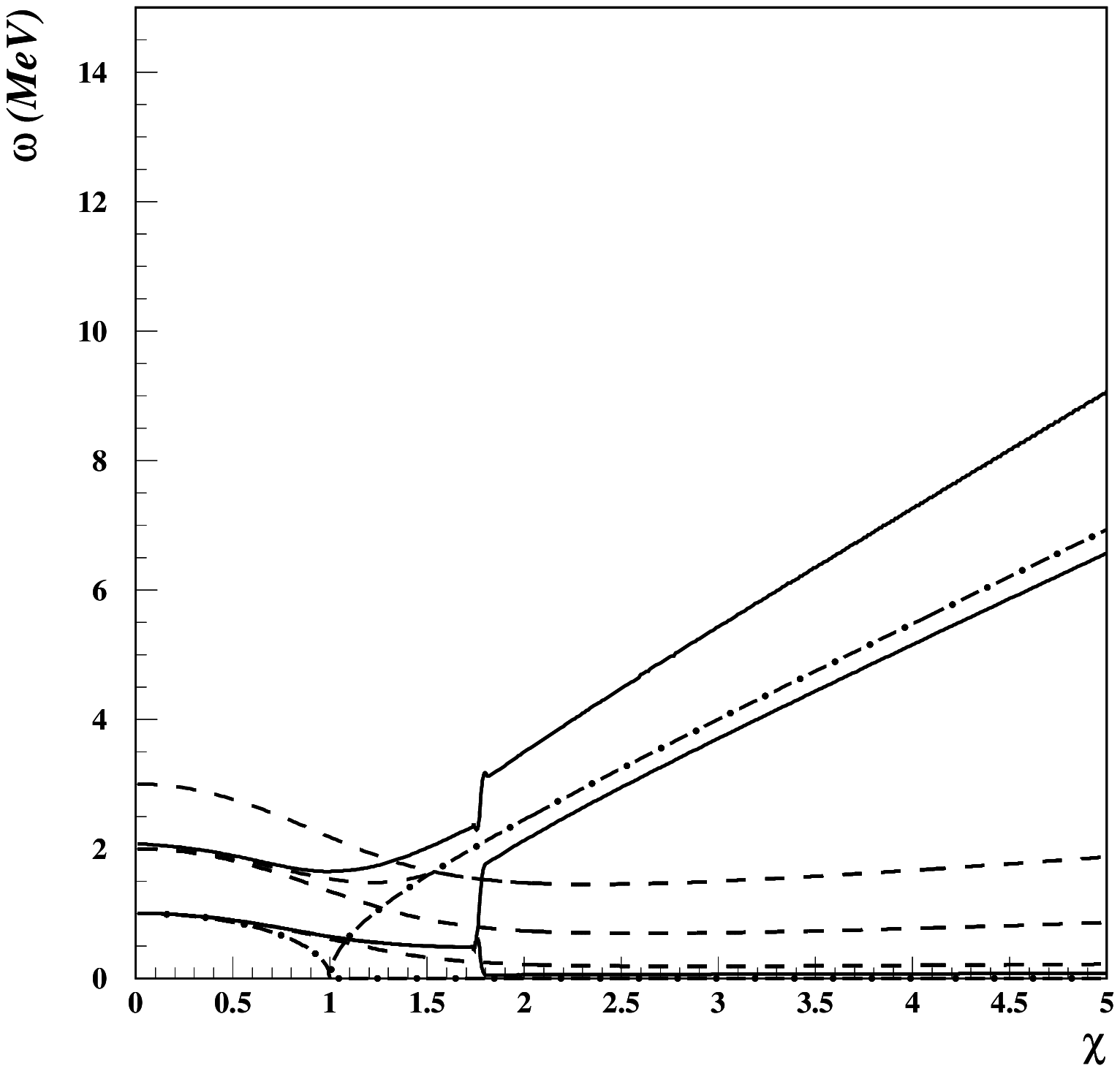}
\vspace{1cm} \caption
{
The same as in Fig. 4, but for $N=20$.
}
\label{fig05}
\end{center}
\end{figure}

\begin{figure}[p]
\begin{center}
\includegraphics[]{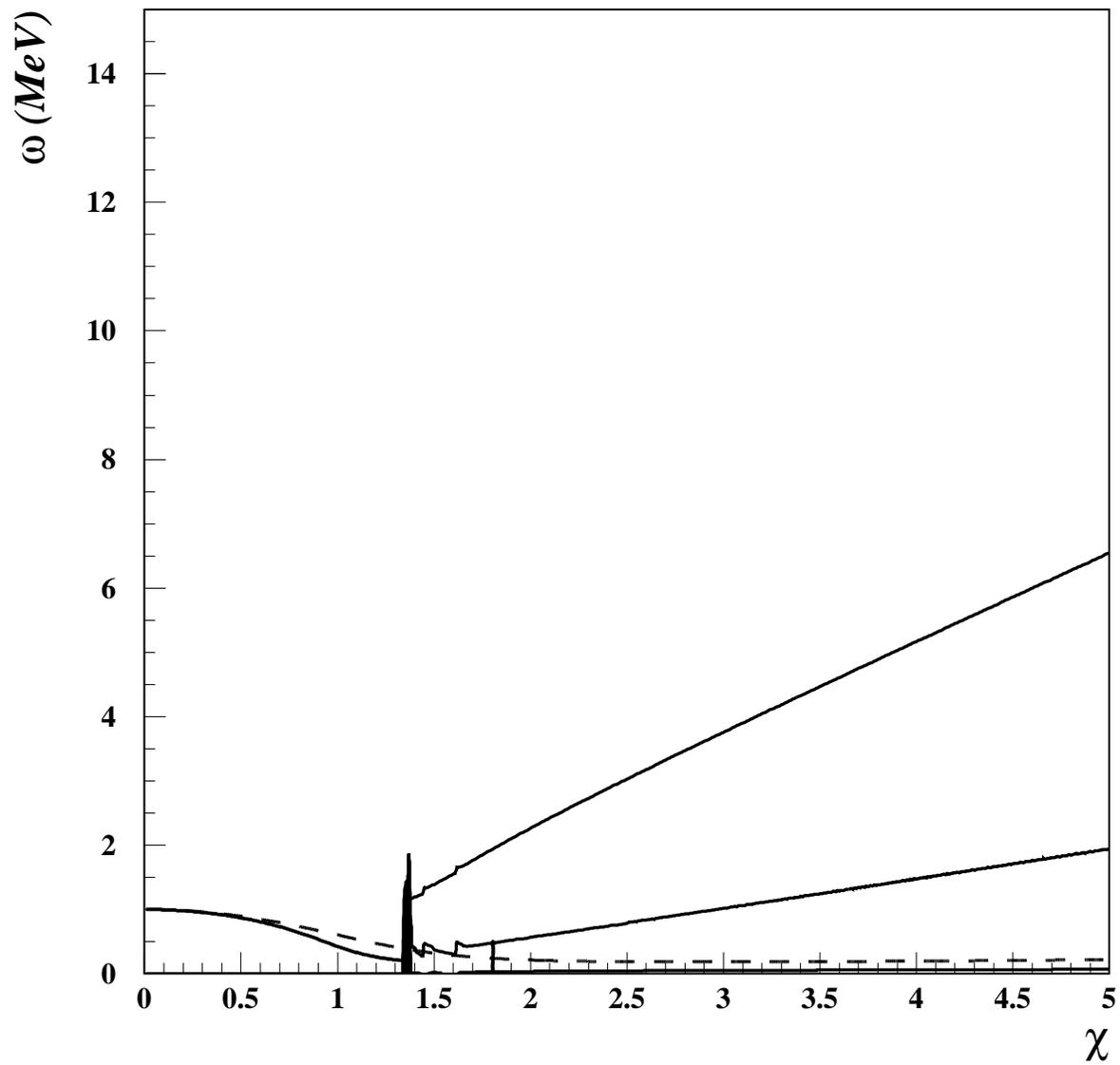}
\vspace{1cm} \caption
{
r-RPA excitation energies versus the strength parameter $\chi$,
for $N=20$, $\Delta\epsilon=0.001~MeV$ (solid lines).
By dashes are given the lowest exact energies.
The Goldstone mode appears at zero energy beyond $\chi\approx 1.4$.
For $\chi\leq 1.4$ appears an unphysical mode of the order
$\Delta\epsilon$ which cannot be distinguished from the abscissa
on the scale of the figure.
}
\label{fig06}
\end{center}
\end{figure}

\begin{figure}[p]
\begin{center}
\includegraphics[]{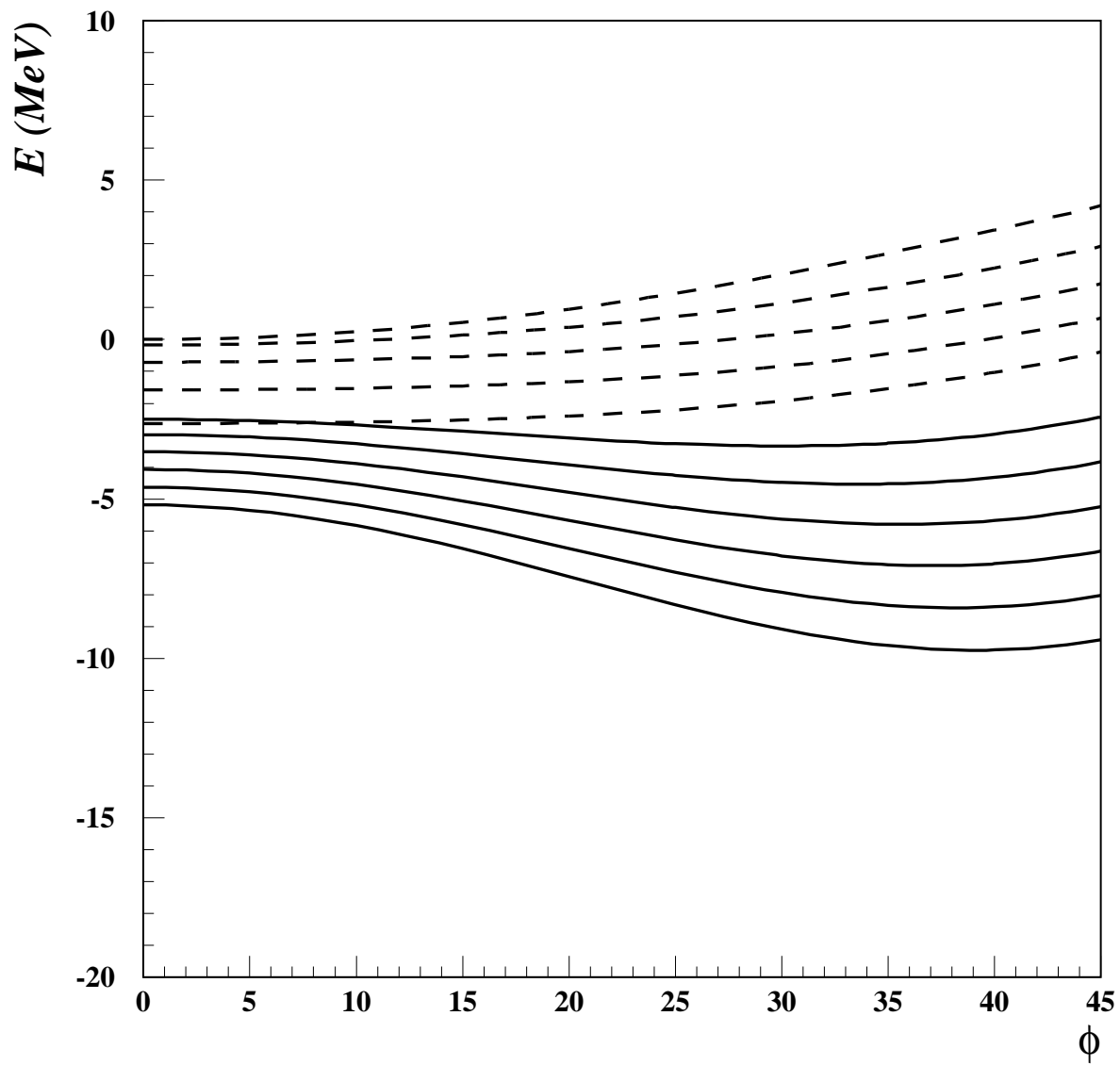}
\vspace{1cm} \caption
{
The SCRPA expectation value of the Hamiltonian versus the angle $\phi$,
for $N=10$ and different values of the strength parameter $\chi$
(from the top of the figure, $\chi=0,0.5,...,5$).
By dashes are given the values for the spherical
region and by solid lines for the deformed region.
}
\label{fig07}
\end{center}
\end{figure}

\begin{figure}[p]
\begin{center}
\includegraphics[]{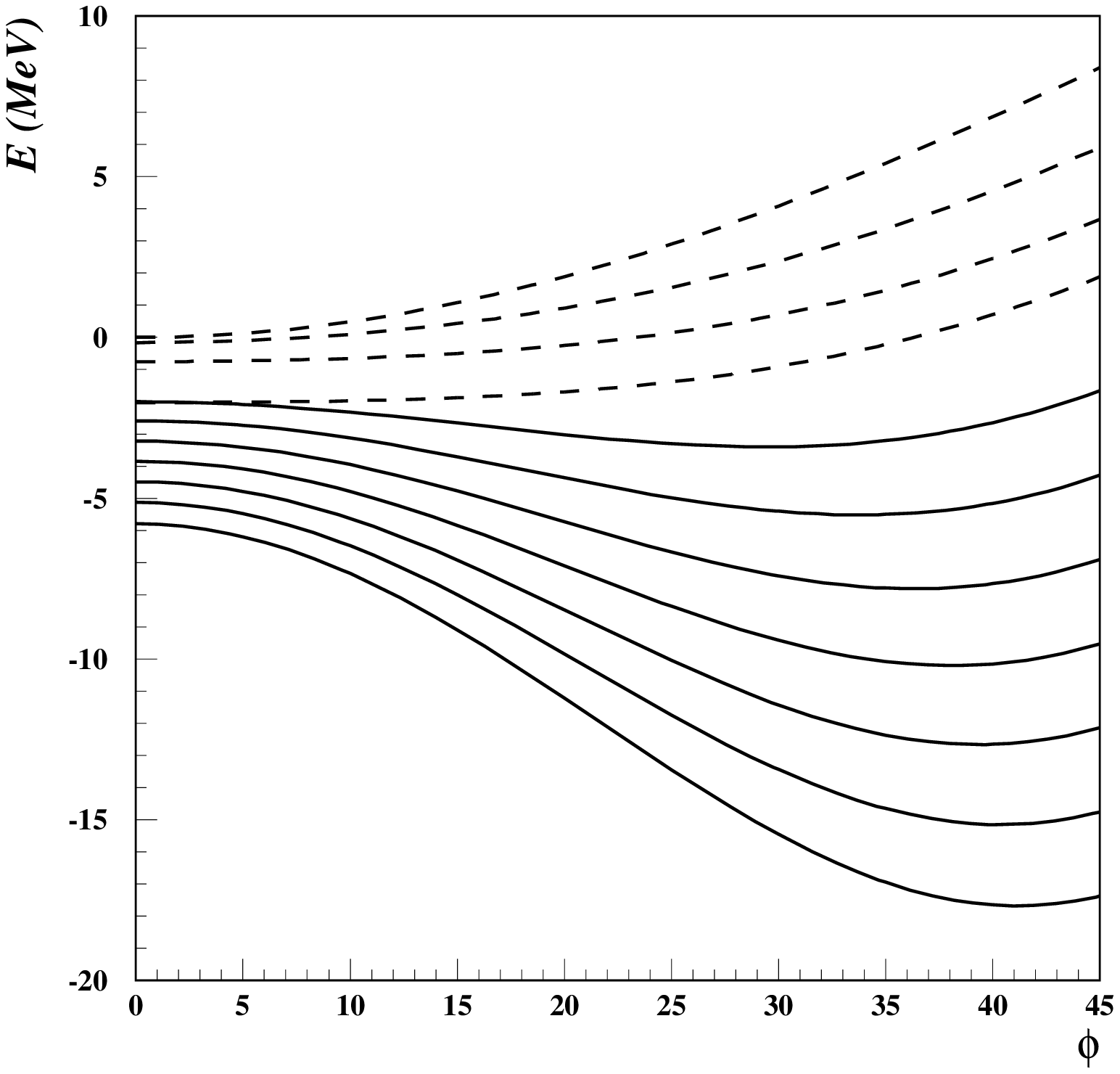}
\vspace{1cm} \caption
{
The same as in Fig. 7, but for $N=20$.
}
\label{fig08}
\end{center}
\end{figure}

\begin{figure}[p]
\begin{center}
\includegraphics[]{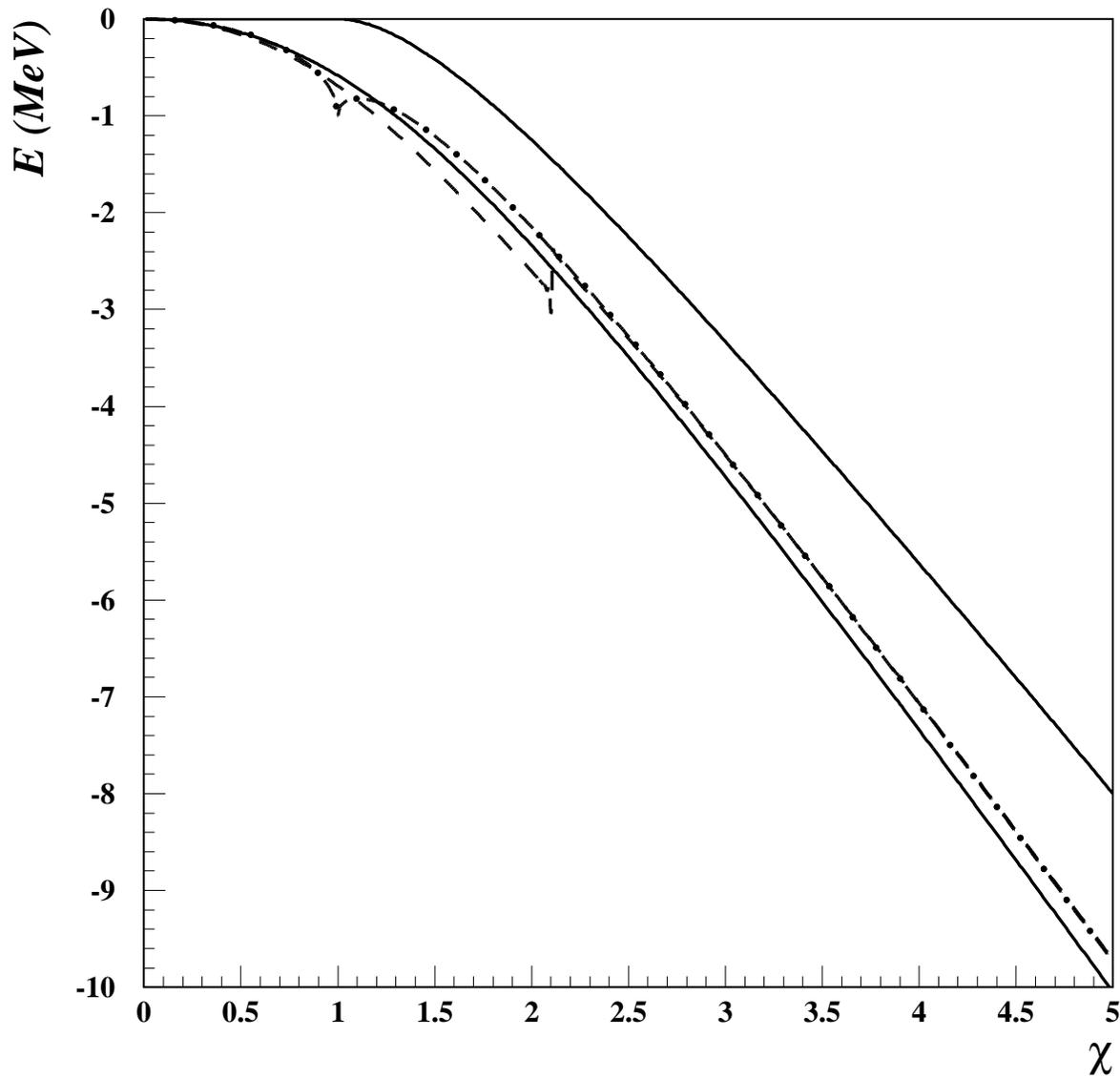}
\vspace{1cm} \caption
{
The expectation value of the Hamiltonian versus $\chi$ for $N=10$.
By the lower solid line are given exact values,
by dashes the SCRPA energies,
by dot-dashes standard RPA results and
by the higher solid line HF values.
Once the SCRPA jumps to the deformed solution at $\chi\approx 2.1$
SCRPA and standard RPA are almost indistinguishible
on the scale of the graph.
}
\label{fig09}
\end{center}
\end{figure}

\begin{figure}[p]
\begin{center}
\includegraphics[]{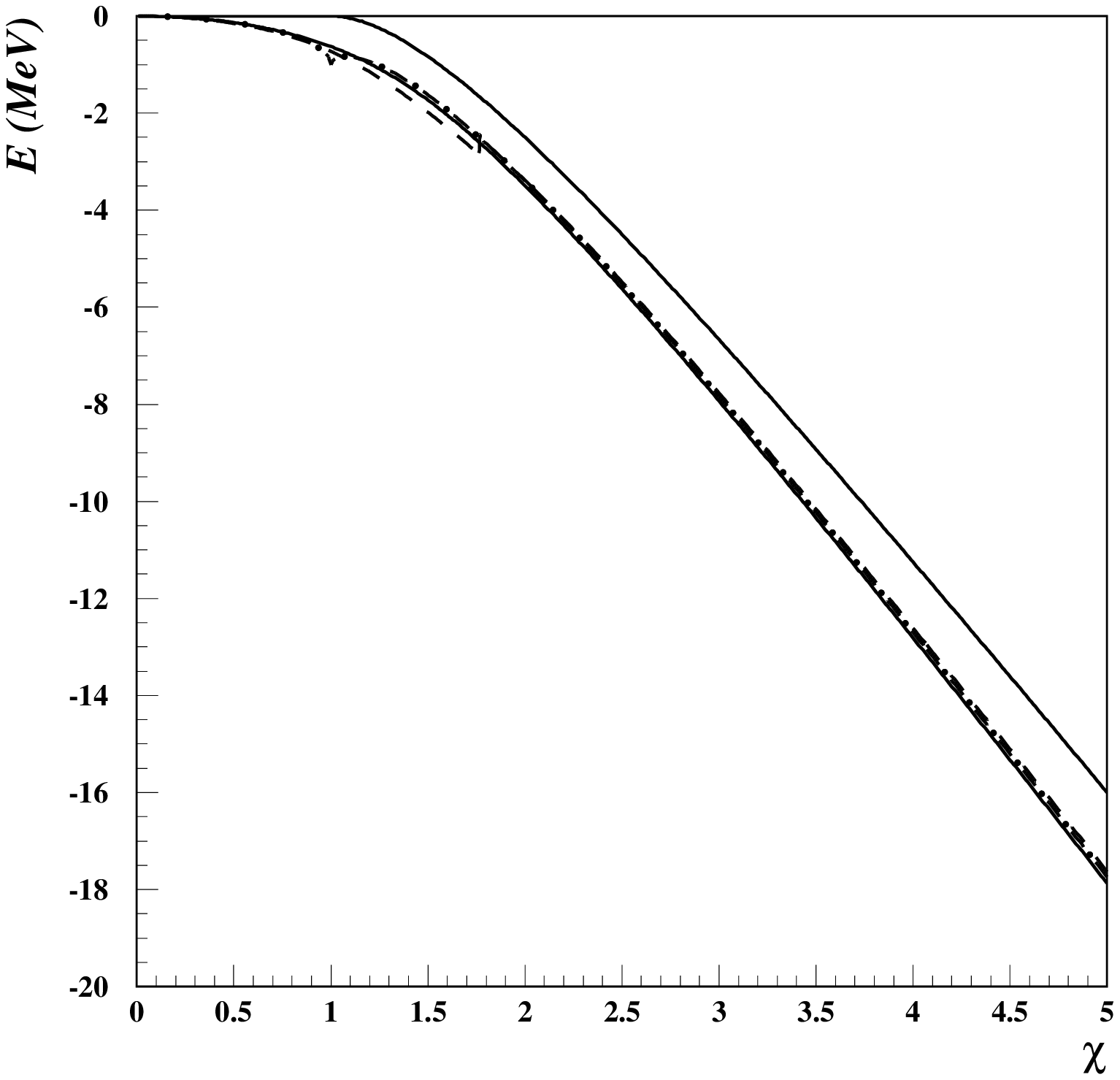}
\vspace{1cm} \caption
{
The same as in Fig. 9, but for $N=20$.
}
\label{fig10}
\end{center}
\end{figure}

\begin{figure}[p]
\begin{center}
\includegraphics[]{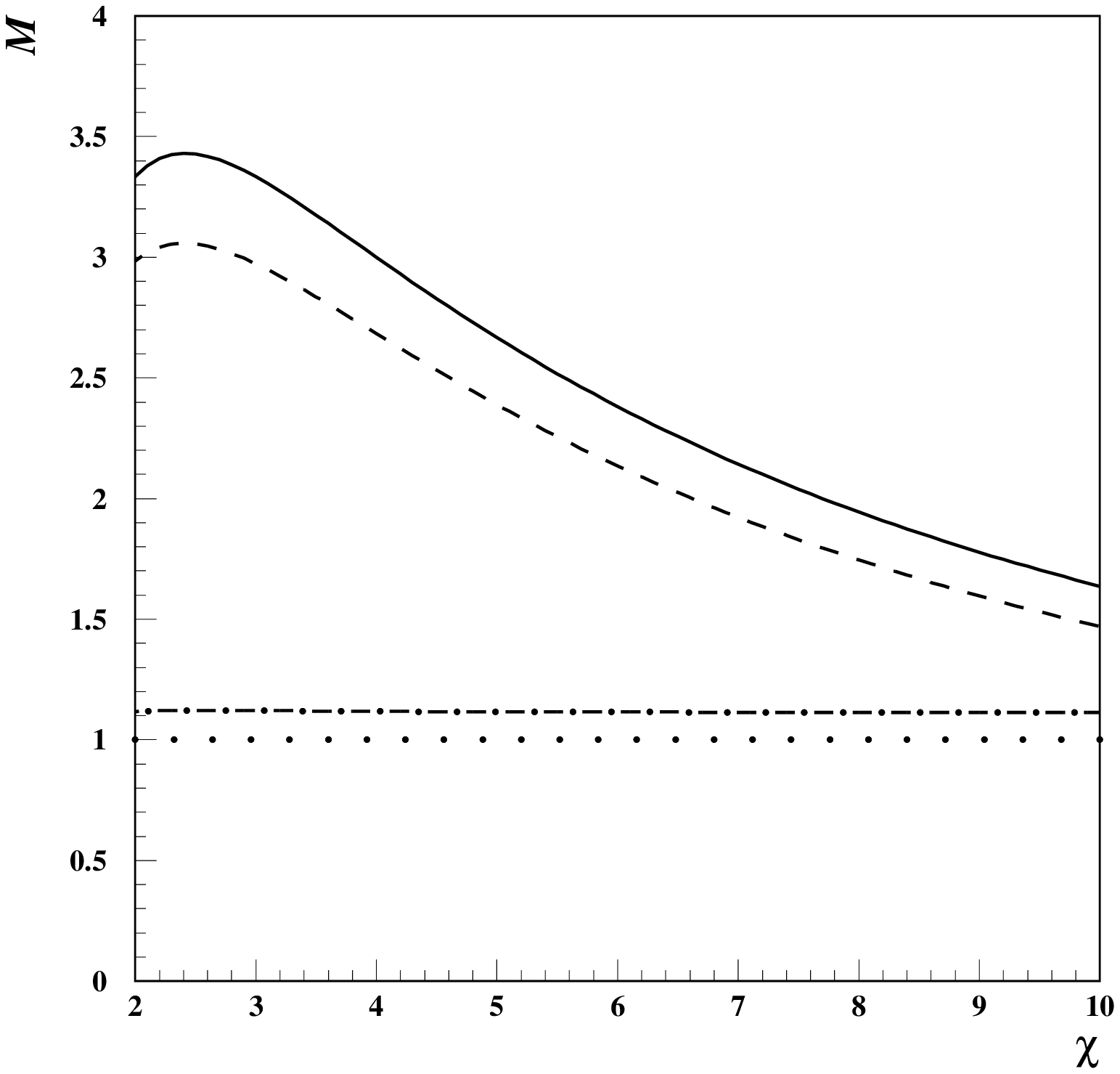}
\vspace{1cm} \caption
{
The standard RPA inertial parameter versus $\chi$ (solid line),
the inertial parameter from the exact energy spectrum (dashed line)
and their ratio (dot-dashed line) for $N=20$.
}
\label{fig11}
\end{center}
\end{figure}

\begin{figure}[p]
\begin{center}
\includegraphics[]{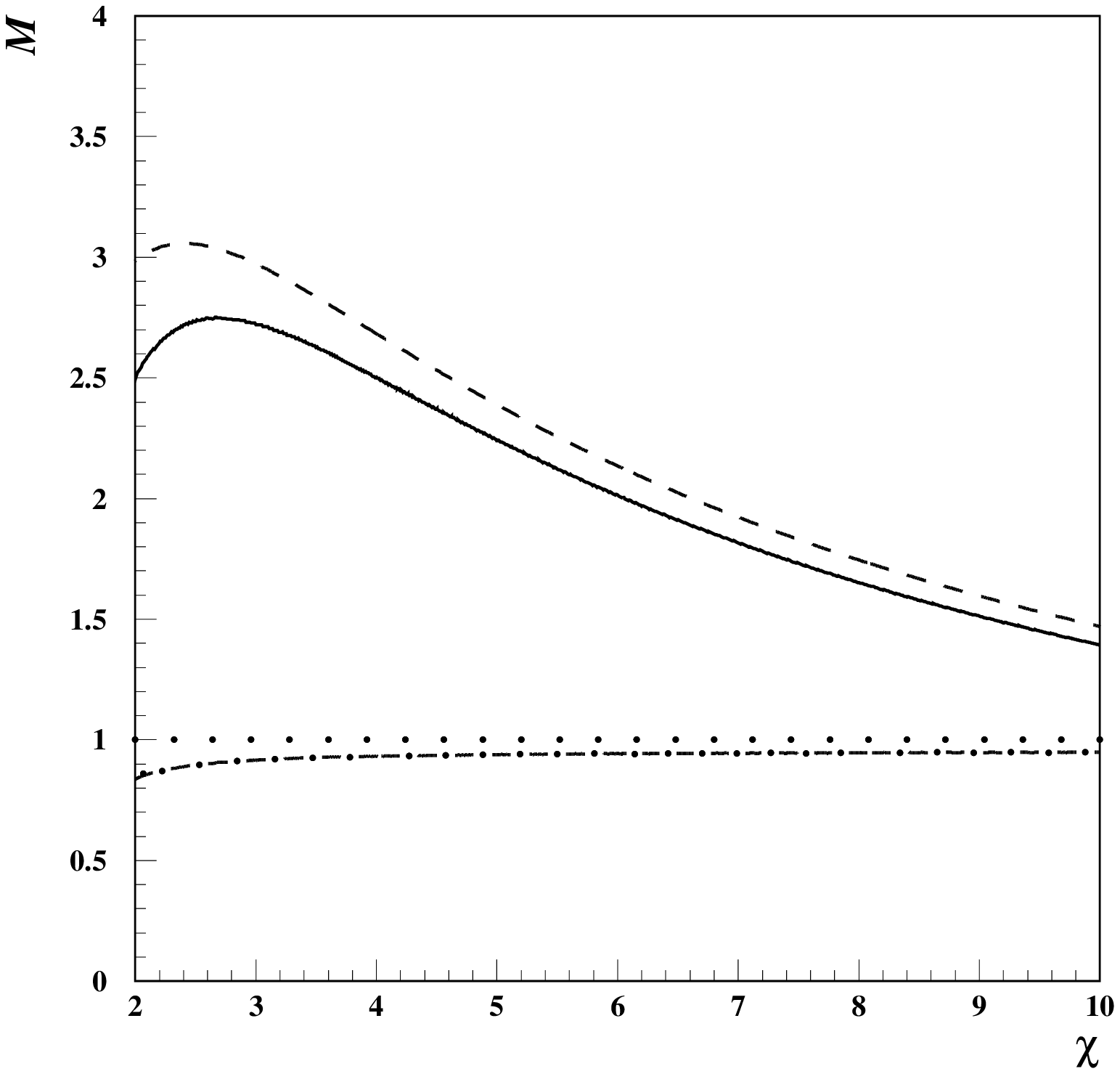}
\vspace{1cm} \caption
{
The SCRPA inertial parameter versus $\chi$ (solid line),
the inertial parameter from the exact energy spectrum (dashed line)
and their ratio (dot-dashed line) for $N=20$.
}
\label{fig12}
\end{center}
\end{figure}

\begin{figure}[p]
\begin{center}
\includegraphics[]{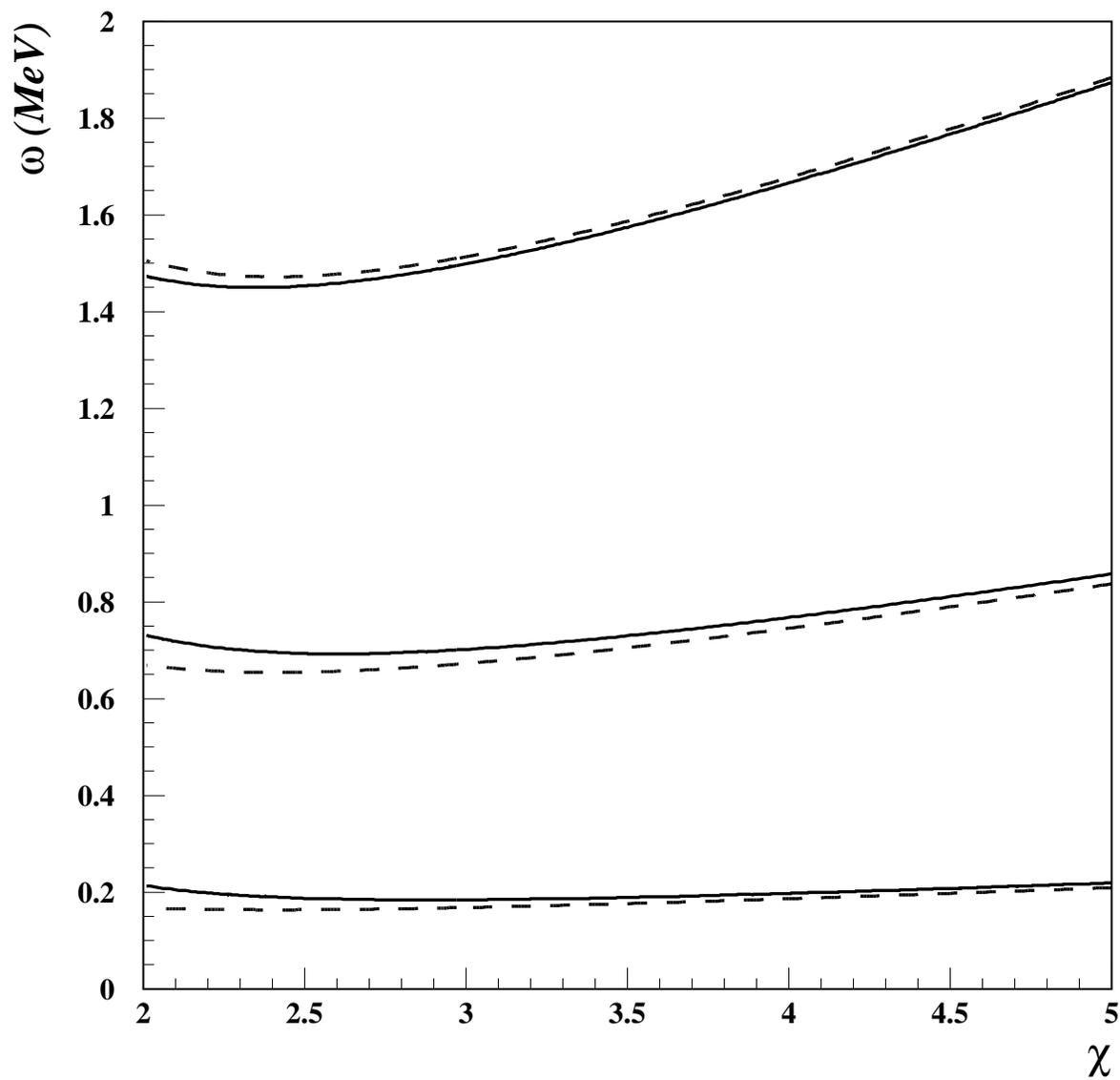}
\vspace{1cm} \caption
{
The SCRPA rotational spectrum (dashed lines)
and the exact energies (solid lines) for $N=20$.
The three levels correspond to $J=1,2,3$.
}
\label{fig13}
\end{center}
\end{figure}

\begin{figure}[p]
\begin{center}
\includegraphics[]{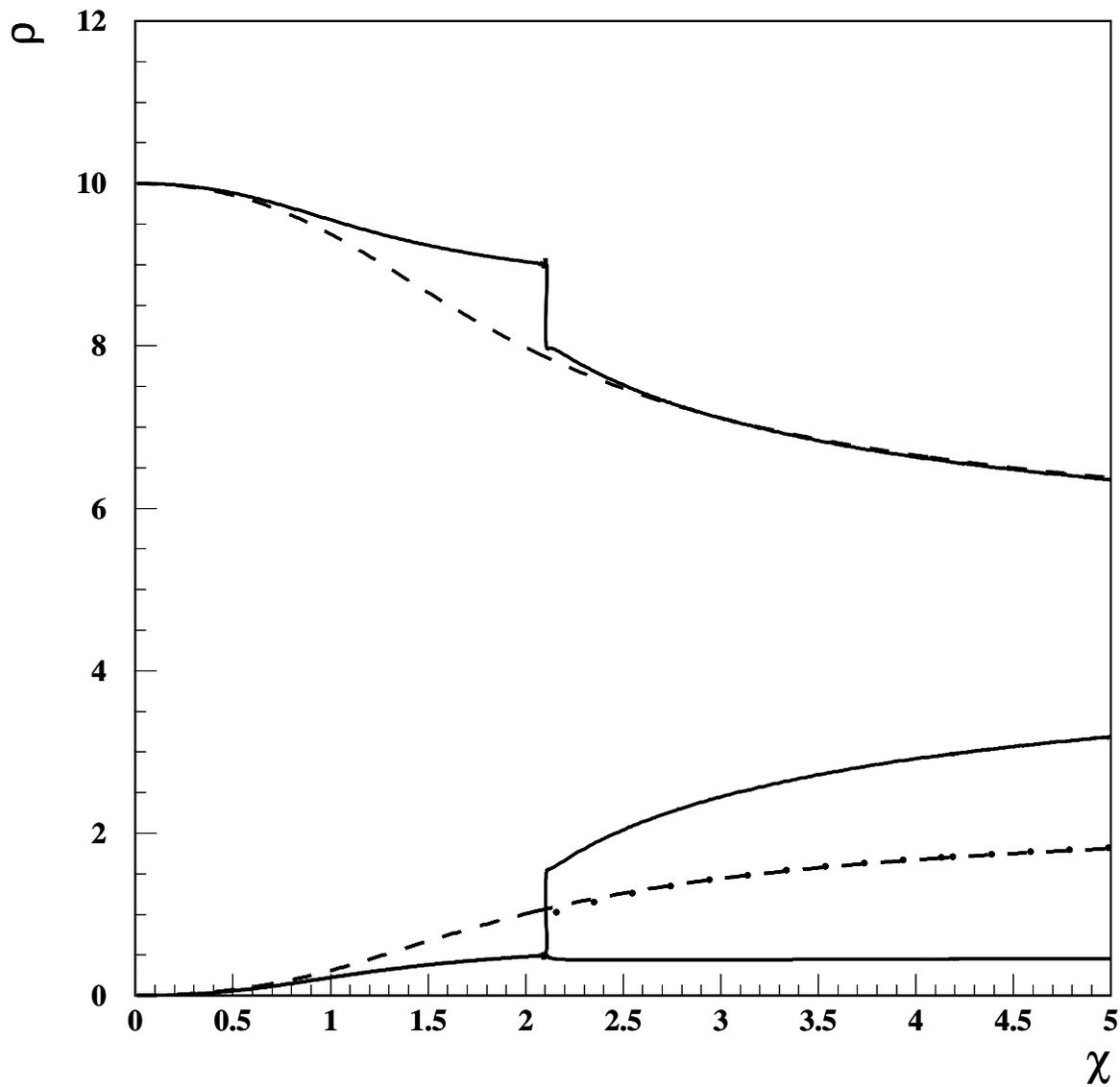}
\vspace{1cm} \caption
{
The one body densities versus $\chi$ for $N=10$.
By solid lines are given SCRPA values and
by dashes the exact results.
The dotted line represents the average of two occupation
numbers of states 1 and 2 for $\chi>\chi_{crt}\approx 2.1$.
}
\label{fig14}
\end{center}
\end{figure}

\begin{figure}[p]
\begin{center}
\includegraphics[]{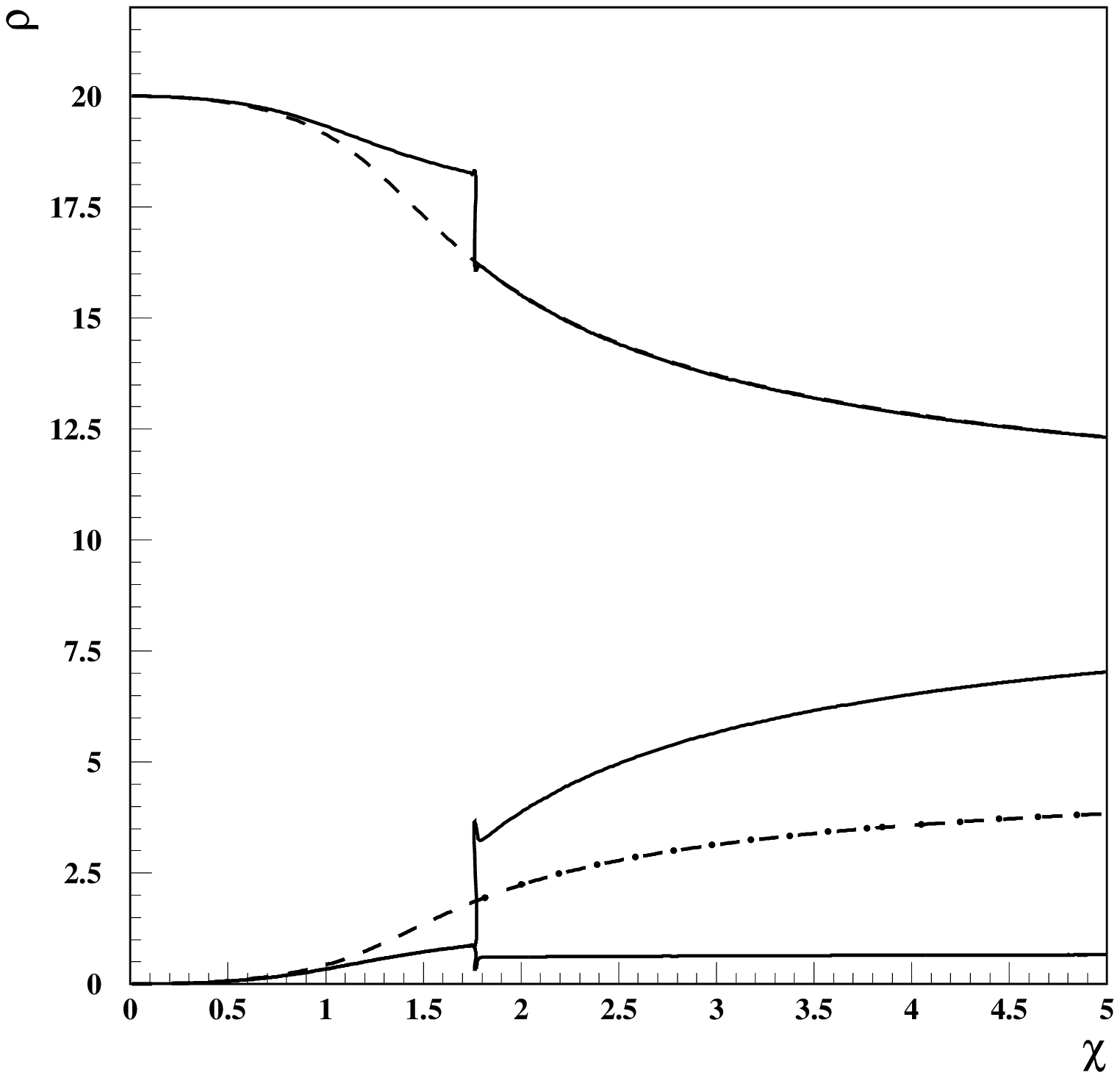}
\vspace{1cm} \caption
{
The same as in Fig. 14, but for $N=20$.
}
\label{fig15}
\end{center}
\end{figure}

\end{document}